\documentclass[1p]{elsarticle}
\usepackage{graphicx,amsmath,amssymb,hyperref}

\title{Arbitrage-free Self-organizing Markets with GARCH Properties:
Generating them in the Lab with a Lattice Model}

\author[fiu,fin]{B.~Dupoyet}
\author[fiu,phy]{H.R.~Fiebig\corref{cor1}}
\author[fiu,phy]{D.P.~Musgrove}

\address[fiu]{Florida International University, Miami, Florida 33199, USA}
\address[fin]{Department of Finance}
\address[phy]{Department of Physics}

\cortext[cor1]{Corresponding author}

\begin{document}

\begin{abstract}
We extend our studies of a quantum field model defined on a lattice having
the dilation group as a local gauge symmetry.
The model is relevant in the cross-disciplinary area of econophysics.
A corresponding proposal by Ilinski aimed
at gauge modeling in non-equilibrium pricing is realized as a numerical
simulation of the one-asset version.
The gauge field background enforces minimal arbitrage, yet allows for statistical
fluctuations. The new feature added to the model is an updating prescription
for the simulation that drives the model market into a self-organized critical state.
Taking advantage of some flexibility of the updating prescription, stylized features
and dynamical behaviors of real-world markets are reproduced in some detail.
\end{abstract}


\maketitle

\section{\label{sec:intro}Introduction}

The analysis and modeling of financial price time series has a long history
\cite{Bachelier:1900,Engle:1982:ARCH,Bollerslev:1986:GARCH}
and has attracted considerable interest at an accelerated pace in the last two decades.
Technological advances have made it possible to collect and process vast amounts of
data. As a result, various stylized facts about the statistics of financial data have been
discovered \cite{Mandelbrot:1963,MantegnaStanley:1995,322494}.
These features are mostly concerned with scaling laws, akin to findings in many systems described
by statistical physics. There, scaling behavior arises from the interaction of many
units in such a way that a critical state is reached. Thus, one may ask if a
financial market, for example, can be modeled based on similar principles.
In generic terms, the building blocks could be many individual agents with suitable
mutual interactions. Indeed, Lux and Marchesi have shown that scaling laws can arise in
such a setting \cite{LuxMarchesi:1999}.

When building a microscopic model it is prudent to rely on a theoretical foundation
supported by evidence. In the present work, we will employ two such principles.
First, arbitrage opportunities will be annihilated during the time evolution of the
market, though admitting statistical fluctuations.
Second, the dynamics of the model will drive it into a self-organized critical state,
thus naturally giving rise to scaling behavior.
Both aspects have been investigated separately in previous work, see \cite{Dupoyet2010107}
and \cite{Dupoyet20113120} respectively.
Here, we merge those elements into a microscopic market model, using numerical
simulation to study its characteristics.

The next section gives an overview of the model's dynamics and definitions.

\section{\label{sec:model}Lattice model}

\begin{figure}[ht]
\center\includegraphics[angle=0,width=100mm]{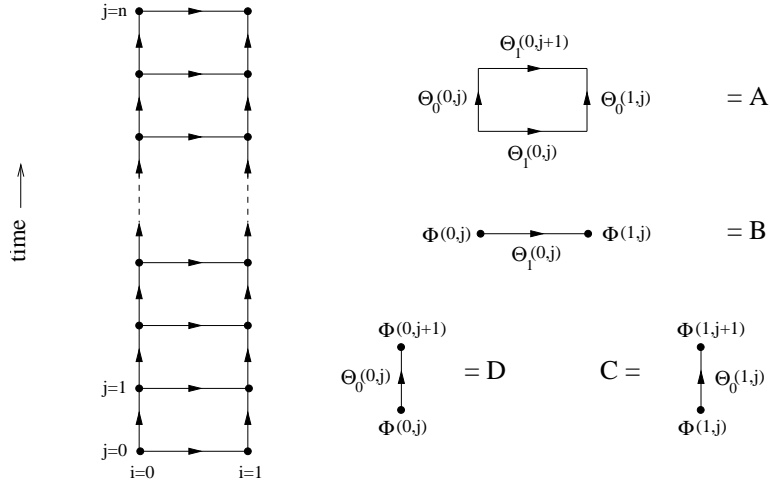}
\caption{\label{fig:gm1lat}Left: Illustration of the ladder geometry of the lattice model and
the label scheme for the fields. Right: Depiction of the gauge invariant elements
A, B, C, D used in the action.}
\end{figure}

Following a proposal by Ilinski \cite{Ilinski:2001:PF} we define a lattice field theory
with a ladder topology as shown in Fig.~\ref{fig:gm1lat}.
In physics terms there are matter fields $\Phi(x)\in{\mathbb R}^+$ defined on
sites $x=(i,j)$, where $j=0\ldots n$ means discretized time, and $i=0,1$ is a spatial index.
These are represented by filled circles in Fig.~\ref{fig:gm1lat}.
In addition, there are gauge fields  $\Theta_\mu(x)\in{\mathbb R}^+$ which live on links
starting at site $x$ in temporal $\mu=0$ or spatial $\mu=1$ direction.
Those are represented by arrowed lines in Fig.~\ref{fig:gm1lat}.

As a model for a financial market, again following \cite{Ilinski:2001:PF}, we interpret
the matter field as instances of an account value, in some unit. At $i=0$ it could be a
cash account, whereas at $i=1$ the value may be interpreted as the number of shares owned
in some financial instrument. The spatial links $\Theta_1(0,j)$, connecting cash and shares,
are simply conversion factors between the corresponding units.
Temporal links $\Theta_0(0,j)$, which connect cash value sites one time step apart, mean
interest rate factors.  
Similarly, temporal links $\Theta_0(1,j)$, starting from a shares site, carry information
about the change in share value one time step apart.

The rationale behind such a model is to describe a market that dynamically evolves
independent of the trading units being used. For example, in comparable markets, the dynamics
should not depend on the specific, notably arbitrary, currency unit being used in transactions.
This, at least, is the hypothesis which should apply to markets trading in like instruments.

Mathematically, this idea is implemented by a quantum field theory with local gauge invariance.
Such has been worked out in great detail in the context of financial
markets \cite{Ilinski:2001:PF}. In a previous work we have studied some aspects of
those ideas using numerical simulation \cite{Dupoyet2010107}. Since the current
work is directly building on the latter, we refer the reader to \cite{Dupoyet2010107}
for the technical details. In particular, we shall use the nomenclature therein.
However, to keep the presentation self contained, the essential building blocks are
discussed in what follows.

The dynamics of the model derives from an action $S[\Theta,\Phi,\bar{\Phi}]$ for the lattice
fields that is invariant with respect to local gauge transformations
\begin{eqnarray}
\Phi(x) &\rightarrow& g(x)\Phi(x) \label{eq01}\\
\bar{\Phi}(x) &\rightarrow& \bar{\Phi}(x) g^{-1}(x) \label{eq02}\\
\Theta_\mu(x) &\rightarrow& g(x) \Theta_\mu(x) g^{-1}(x+e_\mu) \,, \label{eq03}
\end{eqnarray}
where $\bar{\Phi}(x)=1/\Phi(x)$ and $g(x)\in G$ is an element of the dilation
group $G={\mathbb R}^+$, i.e. multiplication by positive real numbers.
Those carry out conversions between (arbitrary) units.
The action is constructed from the elements depicted in Fig.~\ref{fig:gm1lat}.
These are the smallest gauge invariant objects that can be assembled from the fields.

The diagram associated with label A is known as the elementary plaquette
\begin{equation}
P_{\mu\nu}(x) = \Theta_\mu(x) \Theta_\nu(x+e_\mu) \Theta_\mu^{-1}(x+e_\nu) \Theta_\nu^{-1}(x)\,.
\label{eq04}\end{equation}
Its value is interpreted as the gain (or loss) realized through an arbitrage
move \cite{Ilinski:2001:PF}. The global minimum of the classical action $S[\Theta,\Phi,\bar{\Phi}]$
corresponds to zero arbitrage \cite{Dupoyet2010107}.
Quantization of the field is done through the usual path integral formalism.
The partition function thus is defined as the functional integral
\begin{equation}
Z(\beta) = \int[D\Theta][D\Phi] e^{\textstyle -\beta S[\Theta,\Phi,\bar{\Phi}]}\,.
\label{eq05}\end{equation}
In this way stochastic fluctuations about zero arbitrage are allowed.
Their magnitude is regulated by the parameter $\beta$.

Diagrams B,C,D are gauge invariant elements of the form
\begin{equation}
R_\mu(x)=\bar{\Phi}(x)\Theta_\mu(x)\Phi(x+e_\mu)\,,
\label{eq06}\end{equation}
where $e_\mu$ is a unit vector in direction $\mu$.
Diagram C, for example, gives the value of the investment instrument at time $j+1$
divided by its value at $j$, provided we adopt the above interpretation of the fields.
It is a measure for the relative change of the asset value during one time step
\begin{equation}
R_0(1,j)=\bar{\Phi}(1,j)\Theta_0(1,j)\Phi(1,j+1)\,.
\label{eq07}\end{equation}
We also define the related quantity
\begin{equation}
r_{j+1} = \log R_0(1,j)\,,
\label{eq08}\end{equation}
commonly called the `return', indicating a gain ($>0$) or a loss ($<0$) at the
end of the time step. 

\section{\label{sec:update}Updating strategy}

The generation of lattice field configurations as implemented in \cite{Dupoyet2010107}
follows a standard procedure.
Based on the action $S[\Theta,\Phi,\bar{\Phi}]$ the field components are updated through
a heatbath algorithm \cite{Montvay:1994qfl,Creutz:1983}
linked to the partition function (\ref{eq05}).
Periodic boundary conditions (in the time direction) are imposed on all fields as well.
However, in contrast to \cite{Dupoyet2010107}, the updating strategy is modified in
two respects.

First, we do set constraints on the
fields that live on the axis $i=0$, see Fig.~\ref{fig:gm1lat}.
The reasoning here is that we wish to design the model such that the axis
describes a cash account subject to accumulating interest.
The interest rate is endogenously determined. Even at $10\%$ annually the daily
rate factor is $1.0003$ and thus hardly distinguishable from one.
As the model is designed to describe a high frequency market, where the time
extent $n$ translates to typically a day, or so, we wish to set a constraint accordingly.
In a gauge model this is not straightforward, because the meaning of the field components
is gauge dependent. To remedy this situation, gauge fixing is called for.
With reference to (\ref{eq01}-\ref{eq03}) define a gauge transformation
along the axis $i=0$,
\begin{equation}
g(0,j)=\bar{\Phi}(0,j)\,,
\label{eq09}\end{equation}
with $g(x)$ on all other sites being arbitrary.
The gauge transformed fields along the axis, $i=0$, then are
\begin{eqnarray}
\Phi^\prime(0,j) &=& g(0,j)\Phi(0,j) = 1  \label{eq10}\\
\bar{\Phi}^\prime(0,j) &=& \bar{\Phi}(0,j)g^{-1}(0,j) = 1  \label{eq11}\\
\Theta^\prime_0(0,j)&=& \bar{\Phi}(0,j)\Theta_0(0,j)\Phi(0,j+1) = R_0(0,j) \label{eq12}\,.
\end{eqnarray}
In the last equation we recognize the link variable as the (gauge invariant)
return (\ref{eq06}) of the cash holding during one time step. We therefore set
\begin{equation}
\Theta^\prime_0(0,j)=1 \,.
\label{eq13}\end{equation}
In our simulation we choose a random start for the lattice fields.
From there, the constraint $R_0(0,j)=1$ is then implemented by applying the gauge
transformation (\ref{eq09}), and then setting the axis links to one (\ref{eq13}).
During the subsequent updating procedure the axis fields $\Phi(0,j)$ and
$\Theta_0(0,j)$ are never changed.
Nonetheless, the right-hand side of (\ref{eq13}) may differ from
one, depending on the interest rate factor desired.

The next step is to run a heatbath algorithm with the lattice action
$S[\Theta,\Phi,\bar{\Phi}]$ until equilibrium is reached \cite{Dupoyet2010107}.
The lattice field configurations then model a market environment
where arbitrage opportunities exist only briefly, subject to fluctuations due to
the quantum nature of the fields.
In economic terms this model describes an efficient market. Equilibrium, however,
does not seem to be realized in the real world \cite{McCauley:2009}.

Second, subscribing to this paradigm, we introduce a new element which is applied
post equilibrium. In \cite{Dupoyet20113120} we have studied a simple model
where market instances live along a linear chain in time direction.
The sites carry fields $r_j\in{\mathbb R}^+$ which are directly interpreted as returns,
thus having the same meaning as (\ref{eq08}). There is no gauge field in the simple model.
The key ingredient is an updating algorithm that mimics the popular
Bak Sneppen evolutionary model \cite{PhysRevLett.71.4083,PerBak:1996,PhysRevE.53.414}.
The quantity
\begin{equation}
v_j=r_j(r_{j+1}-r_{j-1})
\label{eq14}\end{equation}
turned out to be essential to the field dynamics. In the context of \cite{Dupoyet20113120}
the updating strategy
consists in finding the site $j_s$ for which the absolute value of (\ref{eq14}) is
maximal $|v_{j_s}|=\max\{|v_j|:j=0\ldots n\}$, and then replace $r_{j_s}$ and its
neighbors $r_{j_s\pm 1}$ with random numbers.
Iterating this process leads to a self-organized critical state and produces
price times series, and related statistics, which are almost indistinguishable
from those in real markets \cite{Dupoyet2010107}.

\begin{figure}[ht]
\center\includegraphics[angle=0,width=70mm]{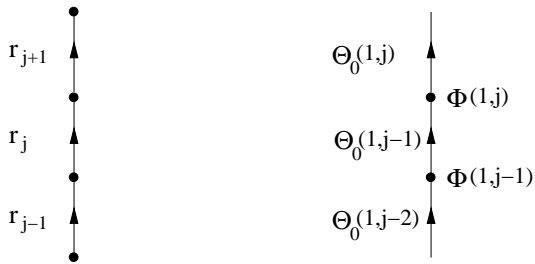}
\caption{\label{fig:gm1bak}Left: Illustration of the returns involved in
the `fitness' criterion (\protect\ref{eq14}) (left), and the field components 
subject to updating (right), done at the `signal' site $j=j_s$.}
\end{figure}

In view of those results it seems desirable to replicate this updating strategy within
the framework of the gauge model as closely as possible. Towards this end, we still do
define the `fitness' measure $v_j$ as in (\ref{eq14}), however, the returns are now given
by (\ref{eq08}), and (\ref{eq07}). Their composition is illustrated in Fig.~\ref{fig:gm1bak}.
The updating prescription proceeds with finding the `signal'
\begin{equation}
V=\max\{|v_j|:j=1\ldots n\}
\label{eq15}\end{equation}
of the field configuration, and the site $j_s$ of its location
\begin{equation}
|v_{j_s}|=V\,.
\label{eq16}\end{equation}
We then update the three field components
$\Phi(1,j_s-1), \Theta_0(1,j_s-1), \Phi(1,j_s)$, which enter the return $r_{j_s}$, and
the two next-neighbor links $\Theta_0(1,j_s), \Theta_0(1,j_s-2)$, see Fig.~\ref{fig:gm1bak}.
Note that $\Phi(1,j_s+1)$ and $\Phi(1,j_s-2)$ are left unchanged. In this way only the three
returns $r_{j_s-1},r_{j_s},r_{j_s+1}$ are affected. This strategy most closely resembles
the updating prescription used in \cite{Dupoyet2010107}.

Updating those field components consists in drawing random numbers from certain
probability distributions. We have chosen those based on the
lattice action $S[\Theta,\Phi,\bar{\Phi}]$ mentioned above.
Heatbath steps using the corresponding Boltzmann-like distribution, see (\ref{eq05}),
are applied to the various field components. Essentially, the probability
distribution for a given field component is given by its local environment.
It is convenient to rewrite the fields as
\begin{equation}
\Theta_\mu(x)=e^{\theta_\mu(x)}\quad\mbox{and}\quad\Phi(x)=e^{\phi(x)}\,.
\label{eq17}\end{equation}
Then, after a gauge transformation, the probability densities for the
gauge fields and the matter fields, respectively, have the form
\begin{eqnarray}
p_\Theta(\theta_\mu(x)) &\propto& \exp(-2\beta\sqrt{L_\Theta\bar{L}_\Theta}\cosh(\theta_\mu(x))\,) \label{eq18} \\
p_\Phi(\phi(x)) &\propto& \exp(-2\beta\sqrt{L_\Phi\bar{L}_\Phi}\cosh(\phi(x))\,) \label{eq19} \,.
\end{eqnarray}
These results are derived in detail in Appendix \ref{sec:A}.
The coefficients $L_\Theta,\bar{L}_\Theta$ and $L_\Phi,\bar{L}_\Phi$ are
independent of $\Theta_\mu(x)$ and $\Phi(x)$, respectively.
The products $L_\Theta\bar{L}_\Theta$ and $L_\Phi\bar{L}_\Phi$ are gauge invariant
and, together with the parameter $\beta$, determine the variance of the probability distributions
for the field components. Those distributions strongly depend on
the local environment at the location of the fields.

Now, at each updating step we randomly draw fields
$\phi^\prime(1,j)$, $j_s-1 \leq j \leq j_s$, from (\ref{eq19}).
Relevant averages considered are
\begin{equation}
a_\theta = \frac{1}{3}\sum_{j=j_s-2}^{j_s}\theta_0(1,j) \quad\mbox{and}\quad
a_\phi = \frac{1}{2}\sum_{j=j_s-1}^{j_s}\phi^\prime(1,j) \,.
\label{eq20}\end{equation}
Updating the fields then is accomplished by replacing
\begin{eqnarray}
\theta_0(1,j)&\longleftarrow&\theta_0(1,j)-\chi a_\theta\,, \quad j_s-2 \leq j \leq j_s \label{eq21}\\
\phi(1,j)&\longleftarrow&\phi^\prime(1,j)-a_\phi\,, \quad j_s-1 \leq j \leq j_s \label{eq22}\,.
\end{eqnarray}
By including the parameter $\chi$ we have introduced a novel feature to the updating process.
While $\chi=1$ essentially mirrors the strategy
in \cite{Dupoyet20113120}, deviations from that value introduce very interesting features to the
model. We will be able to describe a range of different returns distributions and time series,
as will be described in the next section.

Finally, we apply heatbath updates to the two spatial (horizontal) link variables 
$\theta_1(0,j)$, $j_s-1 \leq j \leq j_s$, which connect to the affected matter
fields, see (\ref{eq22}). These links occur in three elementary plaquettes
\begin{equation}
P_{10}(0,j-1) = \Theta_1(0,j-1) \Theta_0(1,j-1) \Theta_1^{-1}(0,j) \Theta_0^{-1}(0,j-1)\,,
\label{eq23}\end{equation}
see (\ref{eq04}), where $j_s-1 \leq j \leq j_s+1$. The reason is that updating $\theta_0(1,j)$,
as prescribed by (\ref{eq21}), changes the plaquettes (\ref{eq23}) and thus upsets the
no-arbitrage environment of the lattice fields. Updating the above links with the
lattice action $S[\Theta,\Phi,\bar{\Phi}]$ rectifies this circumstance.

\section{\label{sec:results}Results}

The simulations were done on a lattice of size $n=782$ with gauge field coupling parameter
$\beta=1$, and the matter field couplings $d_\mu^\pm=\bar{d}_\mu^\pm=1$.
These parameters are the same as in \cite{Dupoyet2010107}, with the one asset model $m=1$.
The number of heatbath update steps was $10^{4}$ to equilibrate the field from a random start.
Final configurations were reached after $4\times 10^{6}$ `signal' updates.

\begin{figure}[ht]
\center\includegraphics[angle=90,width=80mm]{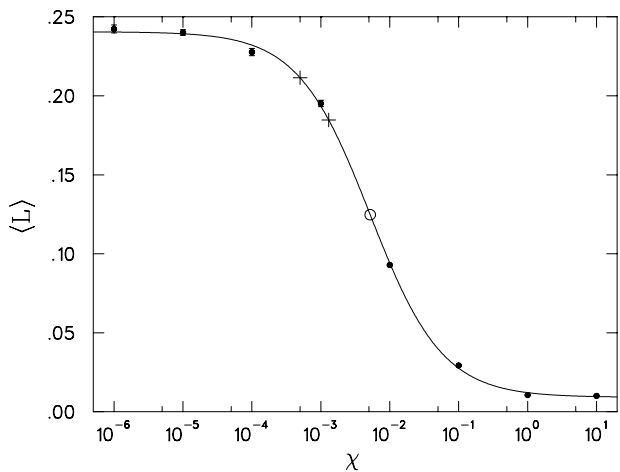}
\caption{\label{fig:lchi}Expectation value $\langle L \rangle$ of the gauge invariant
average link operator (\protect\ref{eq27}) as a function of the update parameter $\chi$
in (\protect\ref{eq21}). The plotting symbols `$\bullet$' and `$+$' correspond to
specific values of $\chi$ for which simulations were made. The symbol `$\circ$' indicates
the symmetry point $\chi=0.0052$ of the fit.}
\end{figure}

First, we discuss the effect of the parameter $\chi$ in (\ref{eq21}). A suitable
observable (order parameter) is the gauge invariant link along the asset axis (\ref{eq07}).
Using the notation (\ref{eq17}) we have
\begin{equation}
R_0(1,j-1)=\exp\left(-\phi(1,j-1)+\theta_0(1,j-1)+\phi(1,j)\right)=\exp(r_j)\,.
\label{eq25}\end{equation}
The updating algorithm, described in Sect.~\ref{sec:update}, employs symmetric probability
distribution functions for $\theta$ and $\phi$. Consequently, the probabilities for
realizing a gain $r_j>0$ and a loss $r_j<0$ are equal. Thus we define the symmetric link
\begin{equation}
L_j = \frac{1}{2}\left(\exp(r_j)+\exp(-r_j)\right)-1=\cosh(r_j)-1
\label{eq26}\end{equation}
and its lattice average
\begin{equation}
L = \frac{1}{n}\sum_{j=1}^{n} L_j\,.
\label{eq27}\end{equation}
Numerically, the value of $L$ is particular to a distinct lattice field configuration. 
We denote the (stochastic) average over field configurations with angle brackets,
here $\langle L \rangle$. In Fig.~\ref{fig:lchi} the dependence
of $\langle L \rangle$ on the parameter $\chi$ is displayed.
The plot symbols `$\bullet$' indicate data points from simulations at
$\chi=10^k,k=-6,-5\ldots +1$. The errors come from $48$ field configurations.
The line curve in Fig.~\ref{fig:lchi} is a four parameter fit, $a_1\ldots a_4$,
to those eight data points with $y=a_1\tanh[a_2(-\log_{10}(x)+a_3)]+a_4$.
Remarkably, there clearly is a transition region in the, approximate,
range $10^{-4} < \chi < 10^{-1}$.
For small values $\chi\ll 10^{-4}$ the average link operator saturates at $0.2406$,
while for large values $10^{-1}\ll\chi$ it tends to $0.0092$, according to the fit.
Using the (crude) conversion $\cosh(r)-1=L$ for simplicity, see (\ref{eq26}), this corresponds to
returns $r\approx\pm 0.68$ and $r\approx\pm 0.14$, respectively, for the above limits of $\chi$.
Those limits may describe valid markets, which could be seen as volatile and 
calm, respectively. In view of this observation the transition region becomes particularly
interesting. It opens up the possibility to simulate markets with a wide range of
features between those extremes. 
Below, we will present results for $\chi=0.0005,0.0013$ within the transition region.
Those are indicated by plot symbols `$+$' in Fig.~\ref{fig:lchi}.

During a simulation, the evolution of the lattice towards a critical state can be monitored,
for example, by
observing the signal $V$, see (\ref{eq15}), as a function of the updating step counter,
say $s=0,1,2\ldots$. Writing $V(s)$ we follow \cite{PhysRevE.53.414} and define
the `gap' function
\begin{equation}
G(x)=\min\{V(s):s\in{\mathbb N}\cup\{0\} \;\mbox{and}\; s\le x \}
\quad\mbox{with}\quad x\in{\mathbb R}^{+}\cup\{0\}\,.
\label{eq28}\end{equation}
This is a decreasing piecewise constant function with discontinuities at certain discrete
values $x_k, k\in{\mathbb N}$.
The set of update steps between $x_{k-1}$ and $x_k$ is called an avalanche of length
$\Lambda_k=x_{k}-x_{k-1}$.
Eventually, as $x\rightarrow\infty$,
the avalanche size diverges and the system has reached criticality \cite{PhysRevE.53.414}.
For an elaboration on these concepts, presented in a context close to the current work,
we refer the reader to \cite{Dupoyet20113120}. We here only show a key result.

\begin{figure}[ht]
\includegraphics[angle=90,width=62mm]{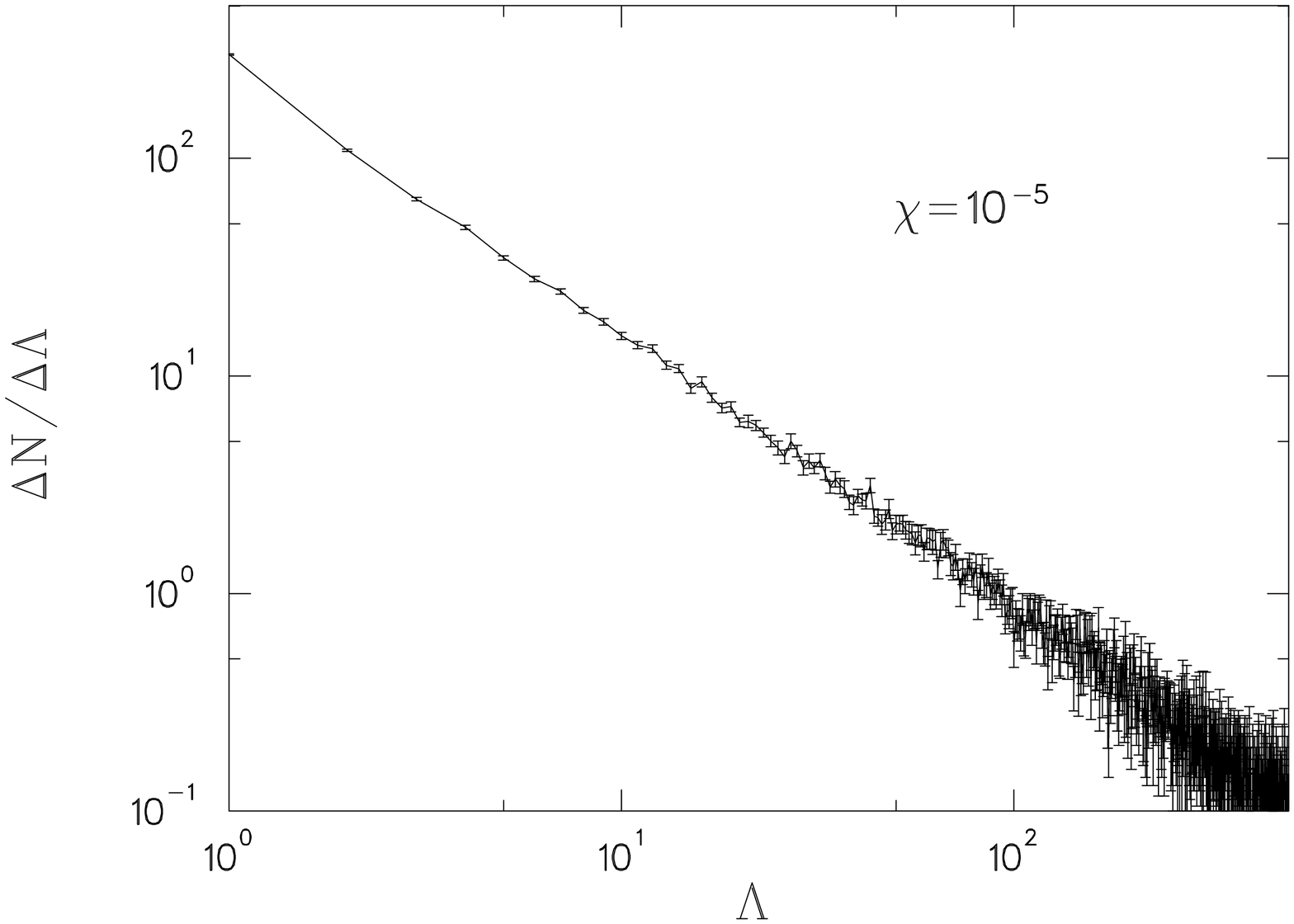}\hspace{\fill}
\includegraphics[angle=90,width=62mm]{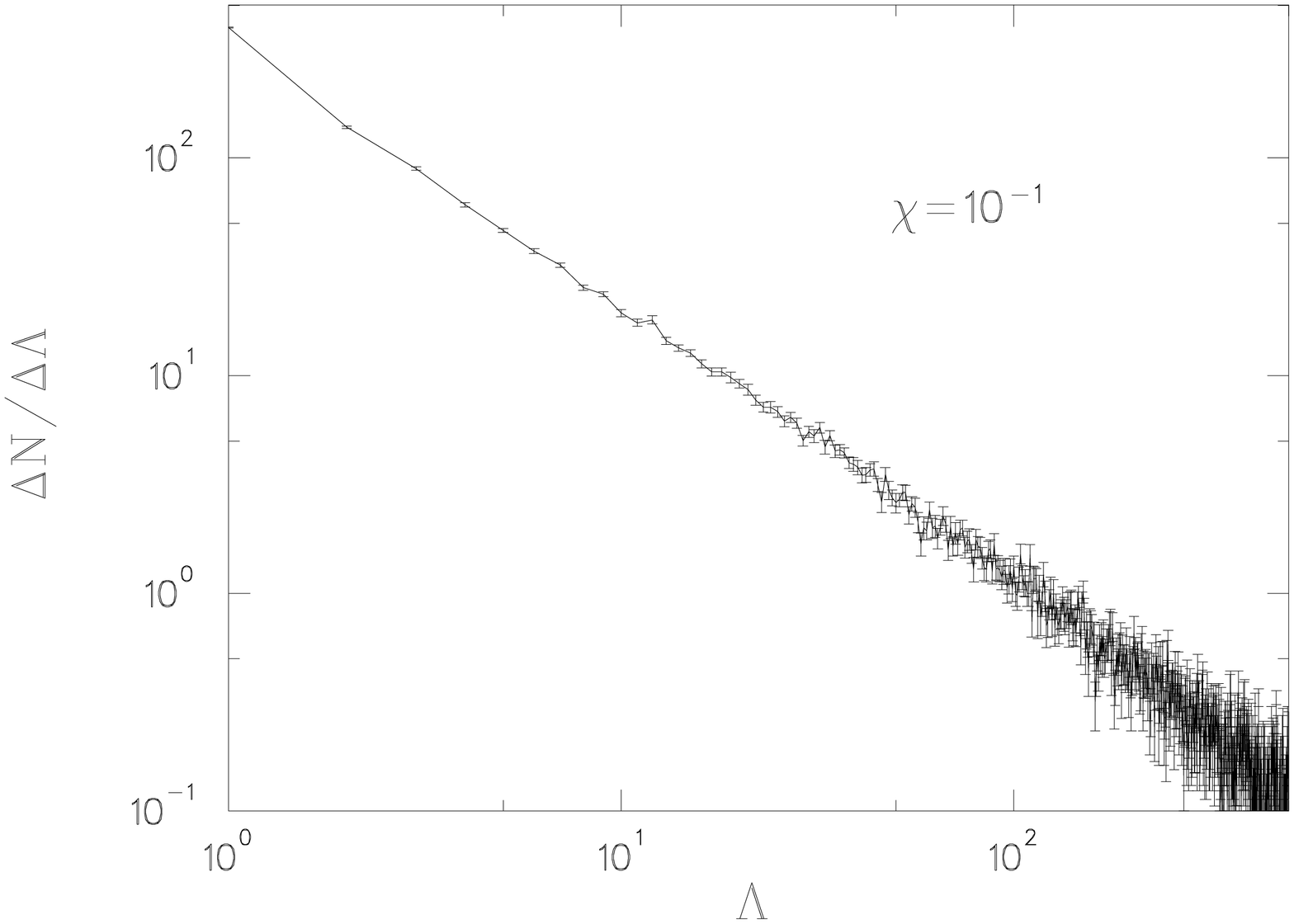}
\caption{\label{fig:ag}Frequency distributions of avalanche sizes for two
values $\chi=10^{-5},10^{-1}$ of the update parameter.}
\end{figure}

In Fig.~\ref{fig:ag} the frequency
distribution $\Delta N/\Delta\Lambda$ of the avalanche sizes is displayed.
Here $\Delta\Lambda$ is a binning interval for the avalanche sizes and
$\Delta N$ is the count of avalanches within that interval. 
We have used 10000 bins with $\Delta\Lambda=1$.
The data points and errors come from an ensemble average over $2400$ independent
lattice simulations with $4\times 10^6$ update steps each.
The $\log$-$\log$ plot clearly shows power law behavior.
A power law indicates scaling, which is a signature feature of a critical system.

For the time being, we continue to present results for the two update parameters
$\chi=10^{-5}$ and $\chi=10^{-1}$. These values correspond to the boundaries of
the transition region, see Fig.~\ref{fig:lchi}. 
Inside that region, the frequency distributions of
avalanche sizes are almost indistinguishable from the results shown in Fig.~\ref{fig:ag}.
Examples of model markets for suitable update parameters in the transition region
are discussed below.

\begin{figure}[ht]
\includegraphics[angle=90,height=45mm]{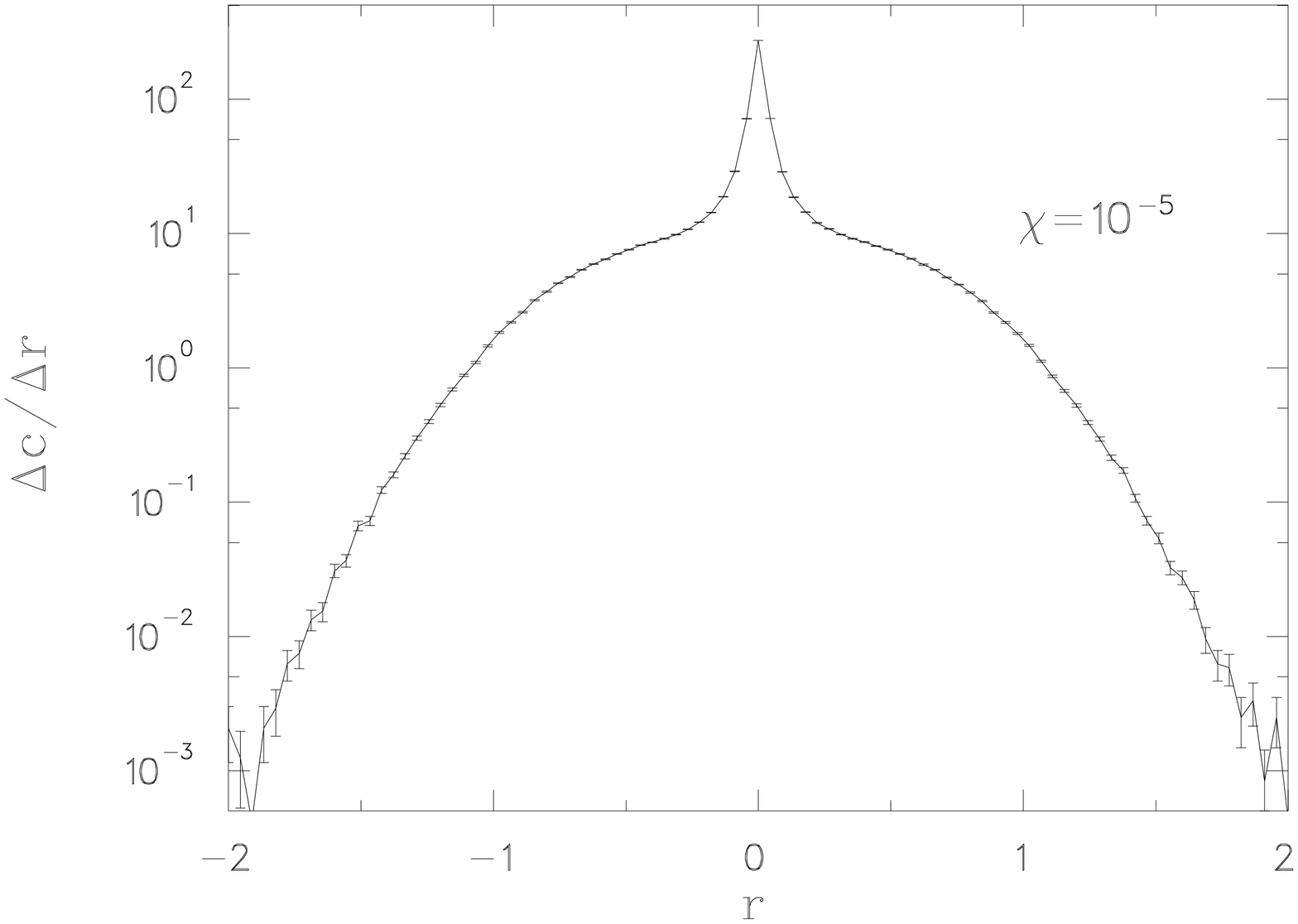}\hspace{\fill}
\includegraphics[angle=90,height=45mm]{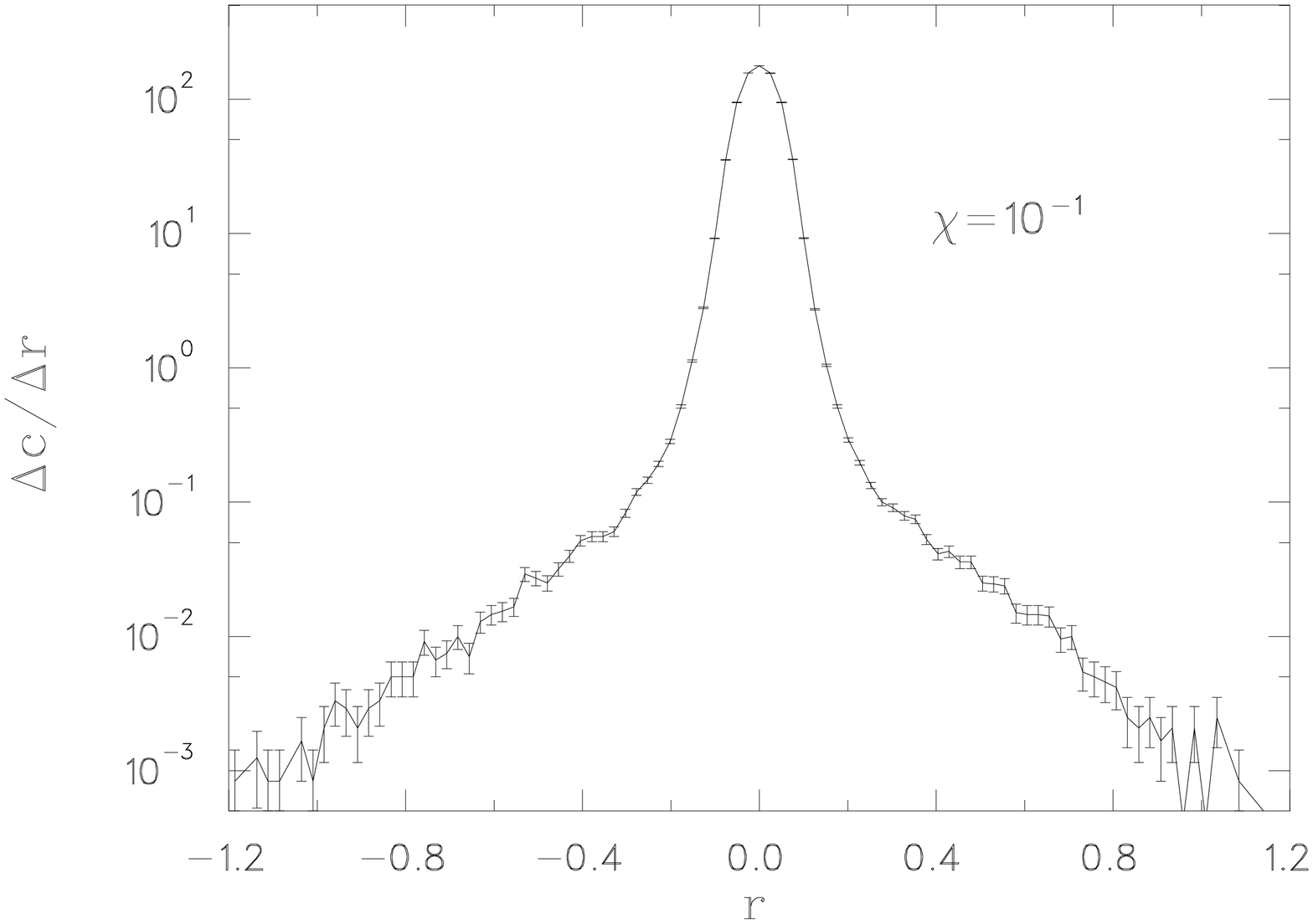}
\caption{\label{fig:gains}Lattice generated gains distributions of the
returns $r$ for the values $\chi=10^{-5},10^{-1}$ of the update parameter.}
\end{figure}

The gains distributions produced by the lattice model
with $\chi=10^{-5},10^{-1}$ are shown in Fig.~\ref{fig:gains}.
The gauge invariant returns $r$, as defined in (\ref{eq08}), are put into bins
of size $\Delta r$, and $\Delta c/\Delta r$ is the number of counts
per bin. The errors are obtained from $2400$ independent simulations.
While both histograms possess fat tails, we observe a distinct difference of
the qualitative features for the distributions. At $\chi=10^{-5}$ a distinctly
pointed central peak sits on very broad bulging tails.
Looking at $\chi=10^{-1}$ the central peak has broadened such that its very top
is almost Gaussian while the tails look linear.
This means that the two distributions cannot be mapped
into each other by simple scale transformations applied to the axes.
The two gains distributions describe genuinely different markets.
Interestingly, this matches our previous assessment of the 
the regions $\chi\ll 10^{-5}$ and $10^{-1}\ll\chi$ discussed in the context
of Fig.~\ref{fig:lchi}.

\begin{figure}[ht]
\includegraphics[angle=90,height=49mm]{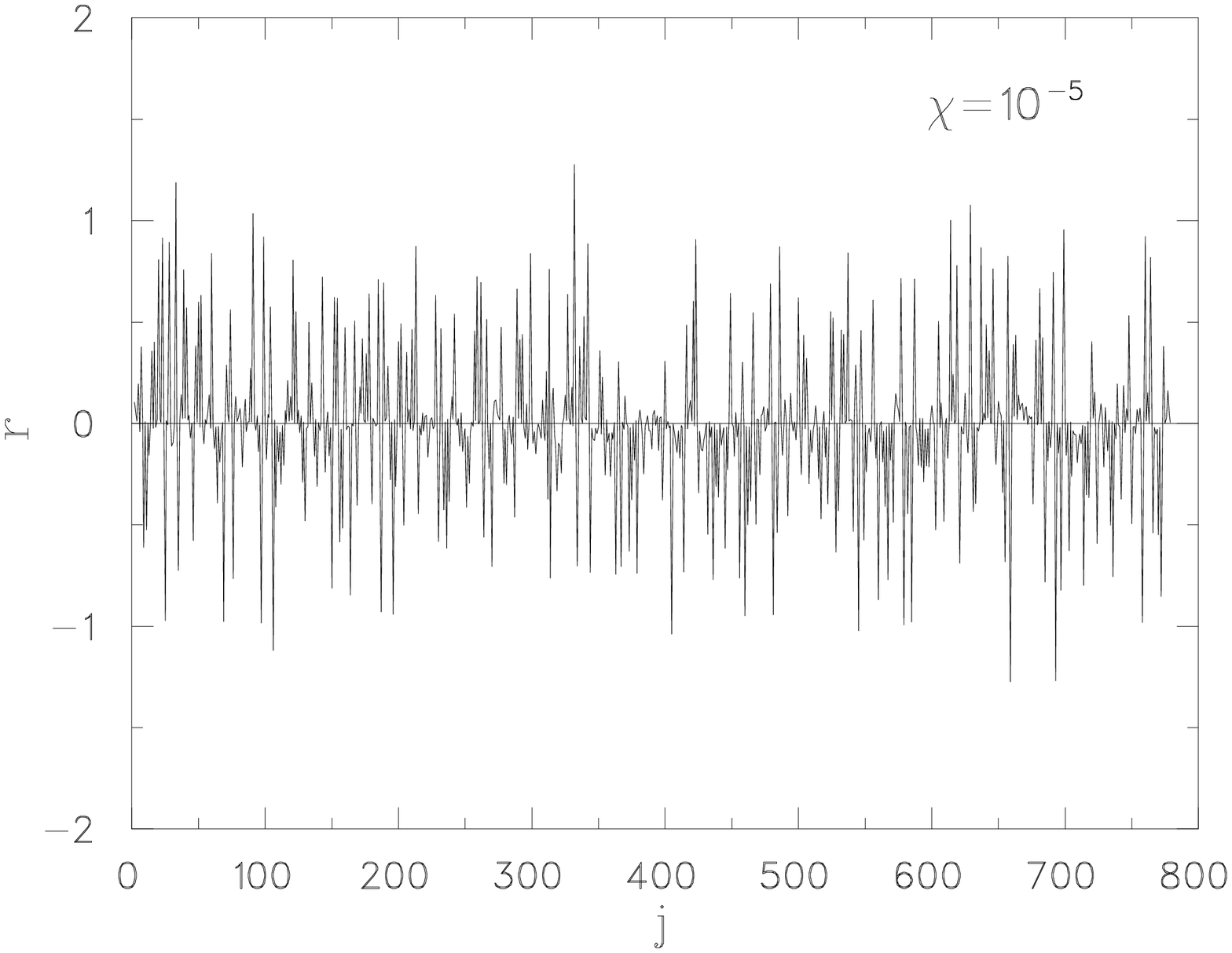}\hfill
\includegraphics[angle=90,height=49mm]{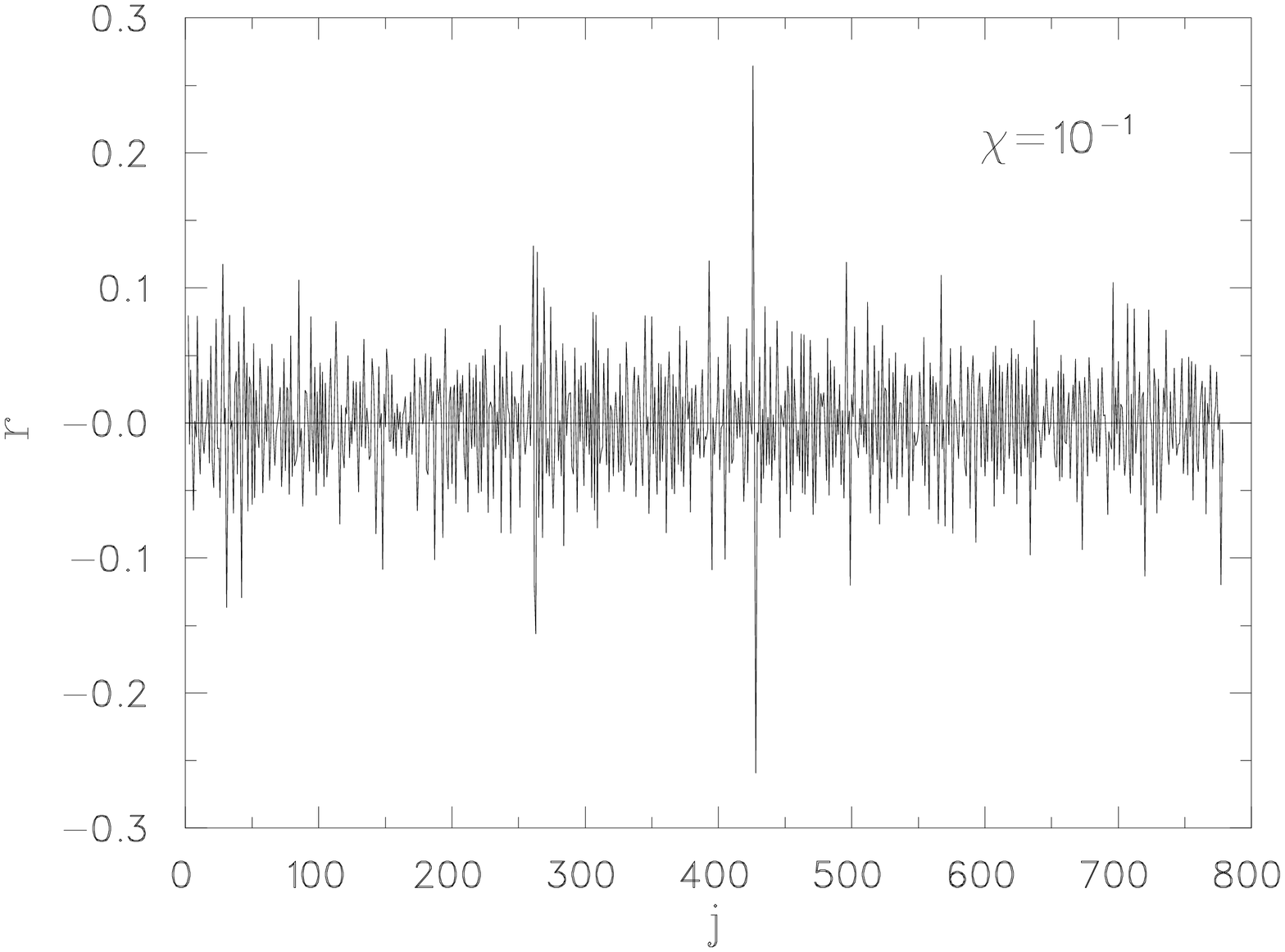}
\caption{\label{fig:tren}Examples of returns time series $r_j$ versus the lattice time $j$
at $\chi=10^{-5}$ and $\chi=10^{-1}$.}
\end{figure}

Examples of returns time series for each of the two
parameters $\chi$ are displayed in Fig.~\ref{fig:tren}.
By visual inspection, both of those exhibit volatility clustering,
yet display different dynamical behavior.
In the realm of $\chi=10^{-5}$ the time series appears to favor one side of the zero 
mark for short periods of time, as compared to the series of $\chi=10^{-1}$
which smoothly fluctuates about zero in either direction. 

For a closer investigation, we have selected the parameters $\chi=0.0005, 0.0013$ from
the transition region. In Fig.~\ref{fig:lchi} their locations are marked by
`$+$' plot symbols. The choice of these values reflects our observation that
model market characteristics, say the gains distribution for example, hardly
change as $\chi$ is decreased from $\approx 0.1$ to $\approx 0.005$. Most of
the market model tuning happens in the upper segment of the 
$\langle L \rangle$ curve in Fig.~\ref{fig:lchi} as $\chi$ further decreases
below $\approx 0.005$.

It is tempting to utilize this range for modeling historical markets of varying
characteristics. However, we leave this to future work. Here, our focus is
on the properties of time series dynamics of the lattice model as being
evaluated with standard financial analysis tools.

\begin{figure}[ht]
\includegraphics[angle=90,height=46mm]{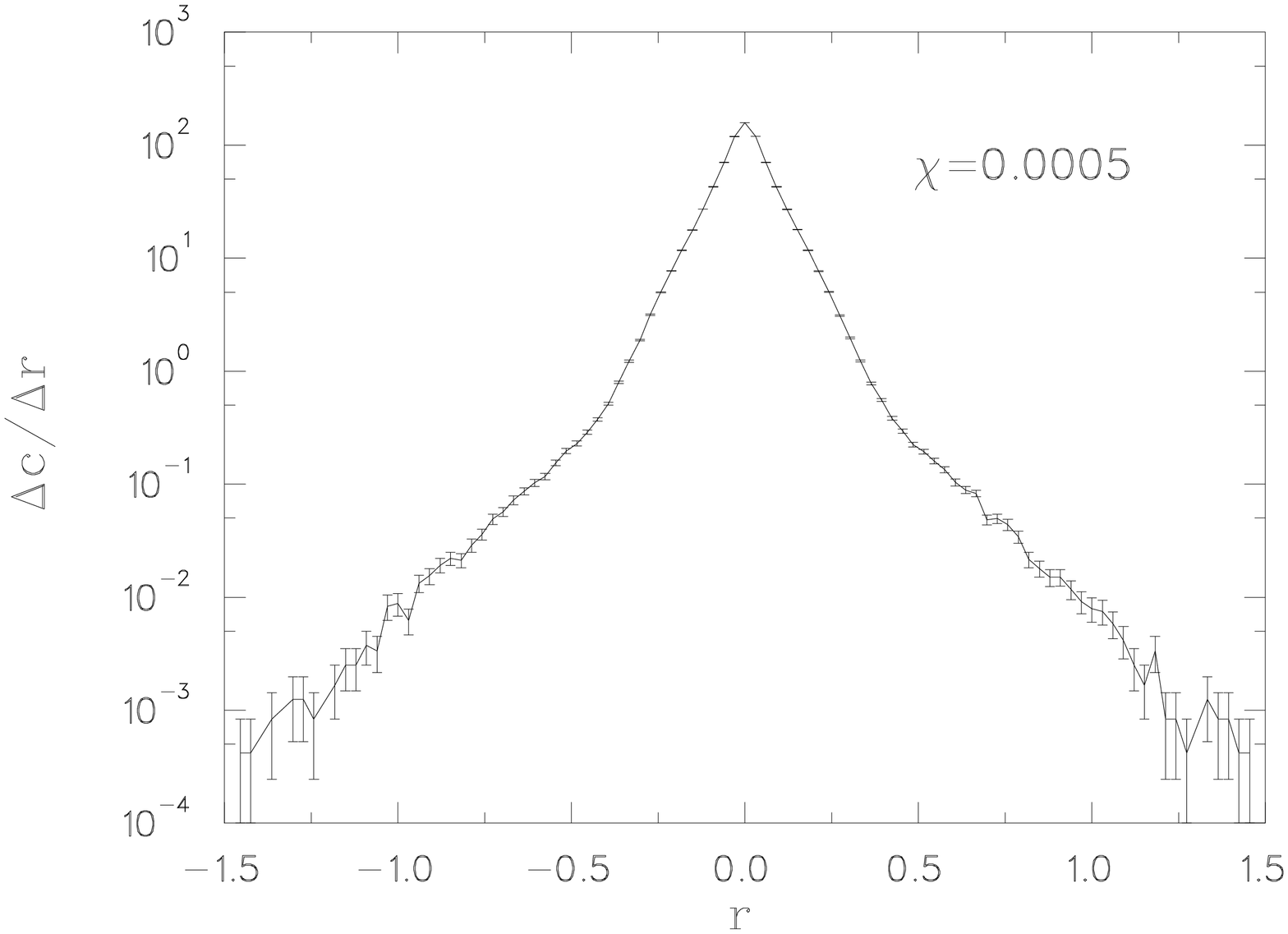}\hfill
\includegraphics[angle=90,height=46mm]{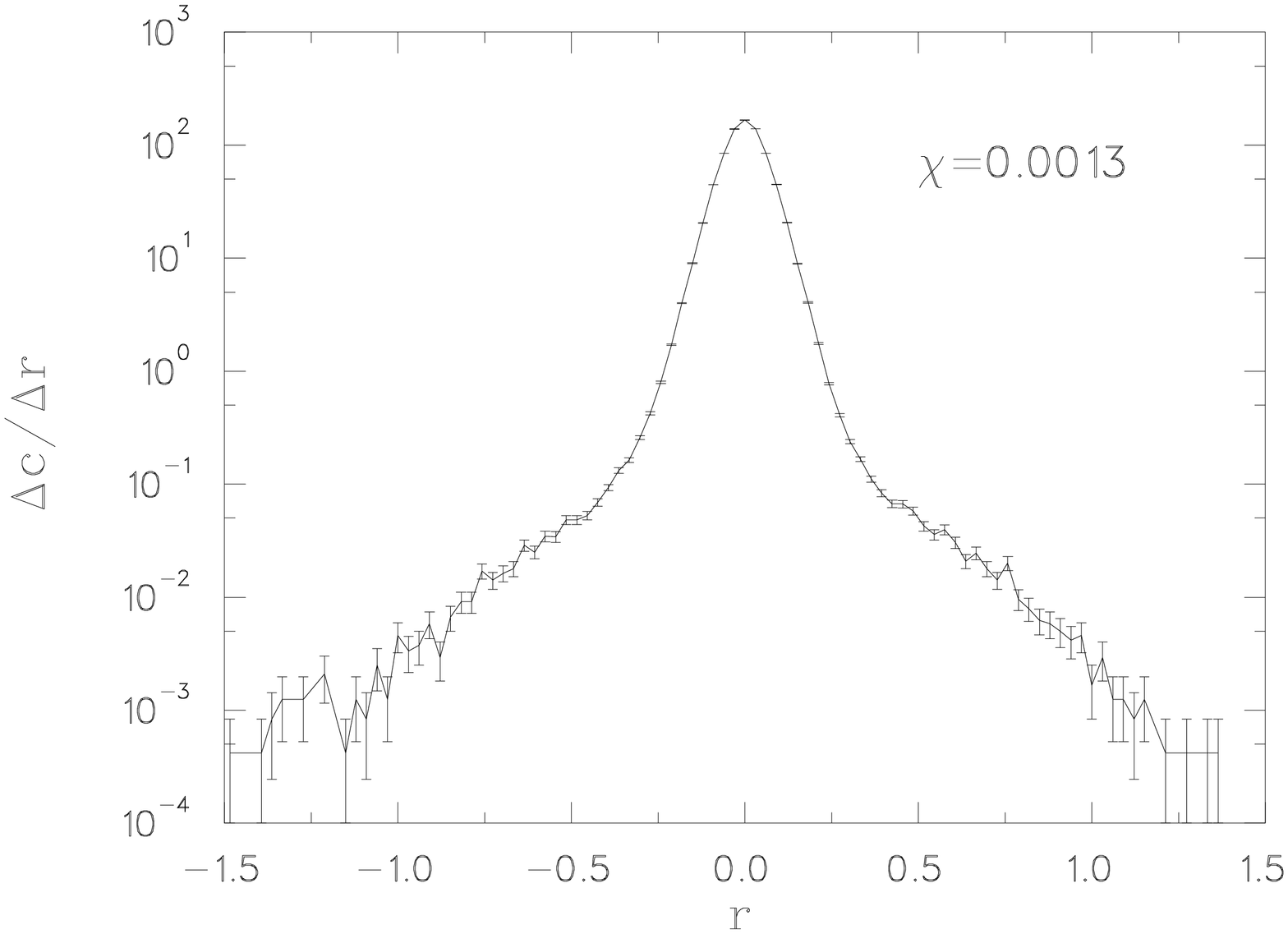}
\caption{\label{fig:gd}Gains distributions from simulations with update parameter
$\chi=0.0005$ and $\chi=0.0013$.}
\end{figure}

The gains distributions for $\chi=0.0005$ and $\chi=0.0013$ are displayed
in Fig.~\ref{fig:gd}. They were obtained from 2400 simulation runs with length $4\times 10^6$ each.
Again, the distributions clearly exhibit fat tails but are otherwise different in their shapes.
The $\chi=0.0005$ distribution puts more emphasis on larger returns than the
$\chi=0.0005$ distribution.
This is echoed in the corresponding returns time series.

\begin{figure}[!ht]
\includegraphics[angle=90,width=134mm]{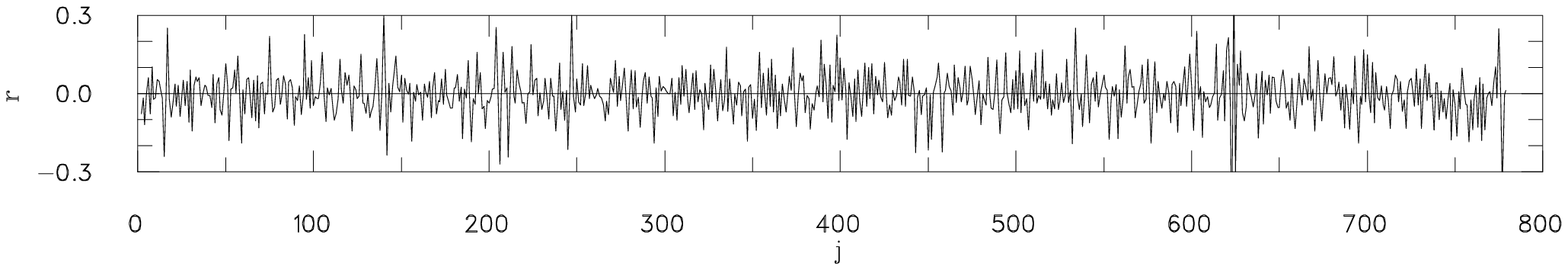}\\
\includegraphics[angle=90,width=134mm]{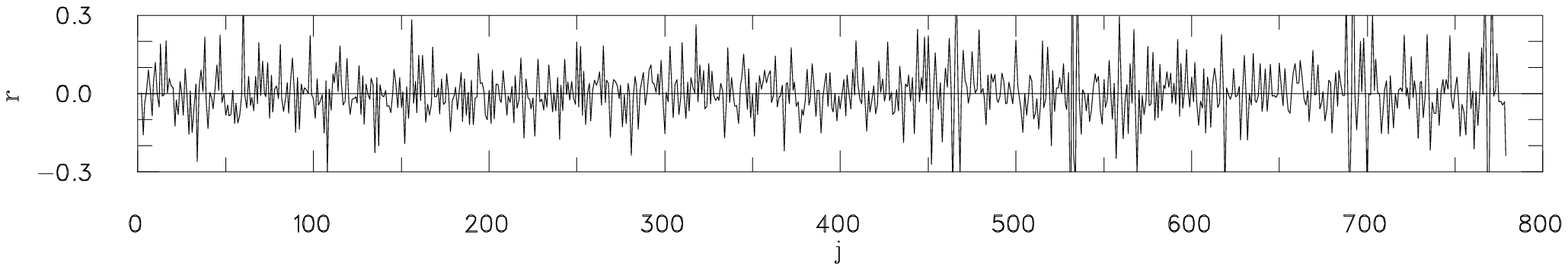}\\
\includegraphics[angle=90,width=134mm]{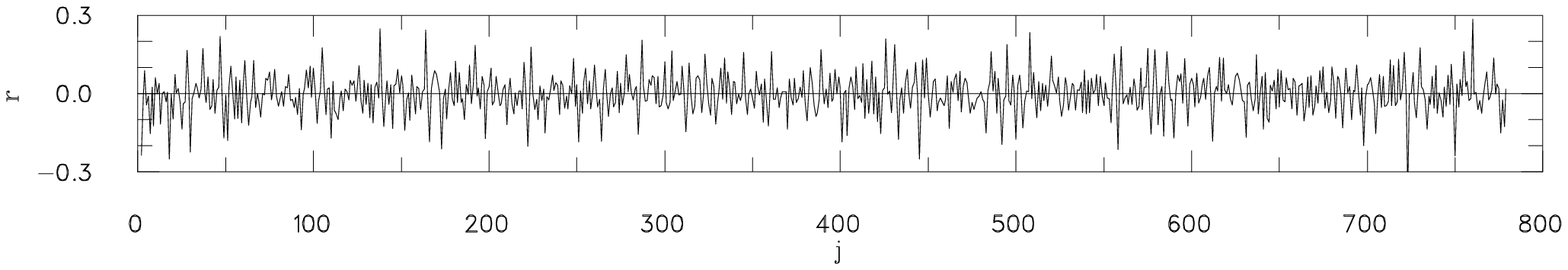}\\
\includegraphics[angle=90,width=134mm]{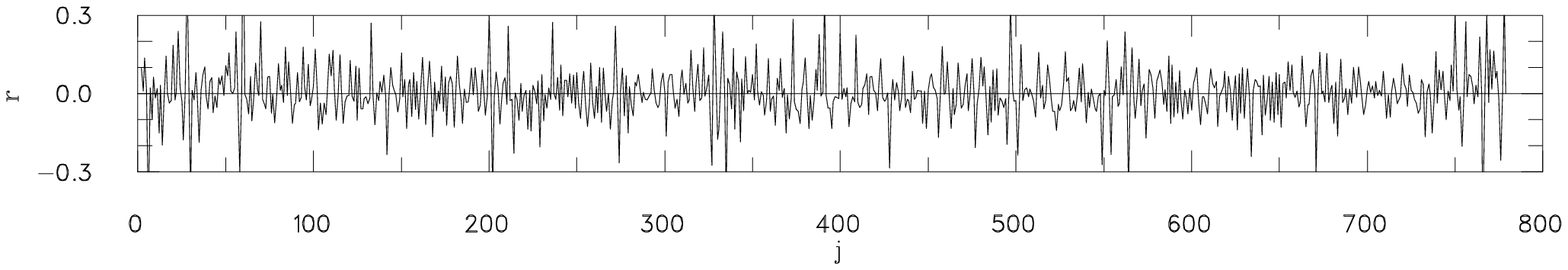}\\
\includegraphics[angle=90,width=134mm]{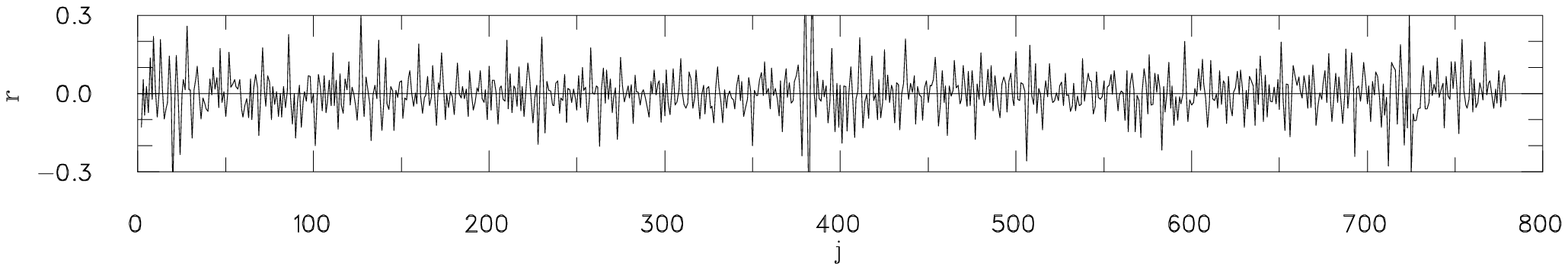}\\
\includegraphics[angle=90,width=134mm]{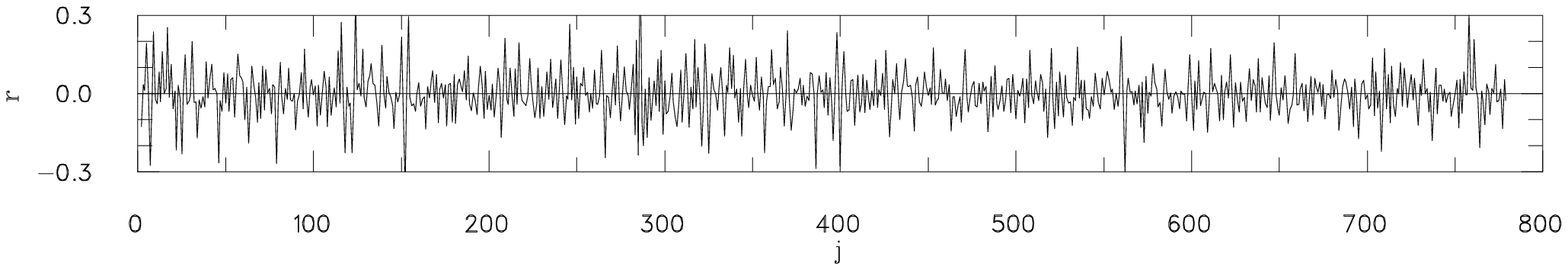}\\
\includegraphics[angle=90,width=134mm]{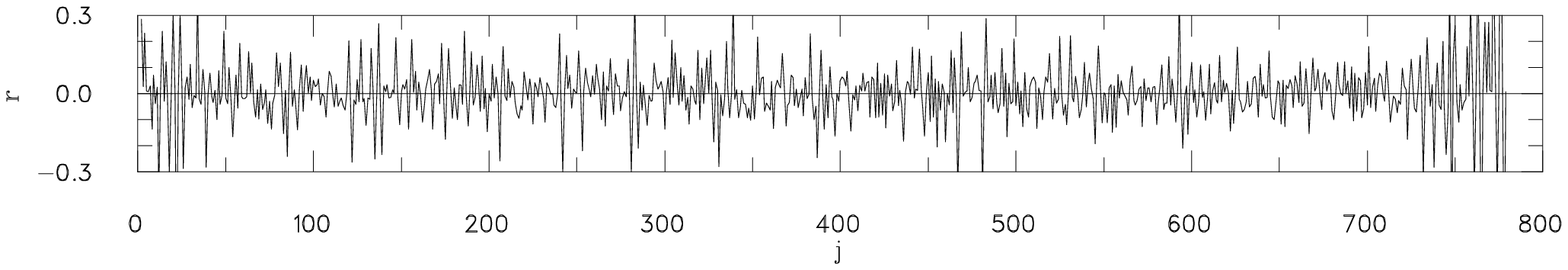}\\
\includegraphics[angle=90,width=134mm]{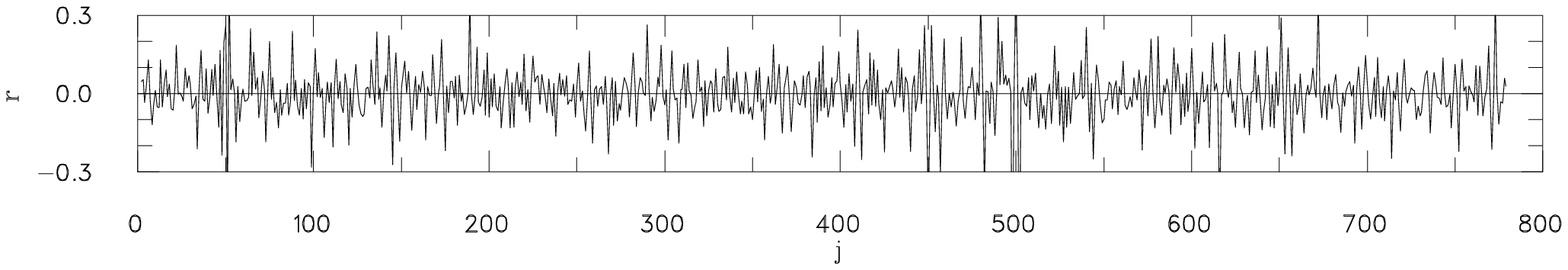}
\caption{\label{fig:tr05}Samples of returns time series for the update parameter
$\chi=0005$. The corresponding gains distribution is shown
in Fig.~\protect\ref{fig:gd}.}
\end{figure}

\begin{figure}[!ht]
\includegraphics[angle=90,width=134mm]{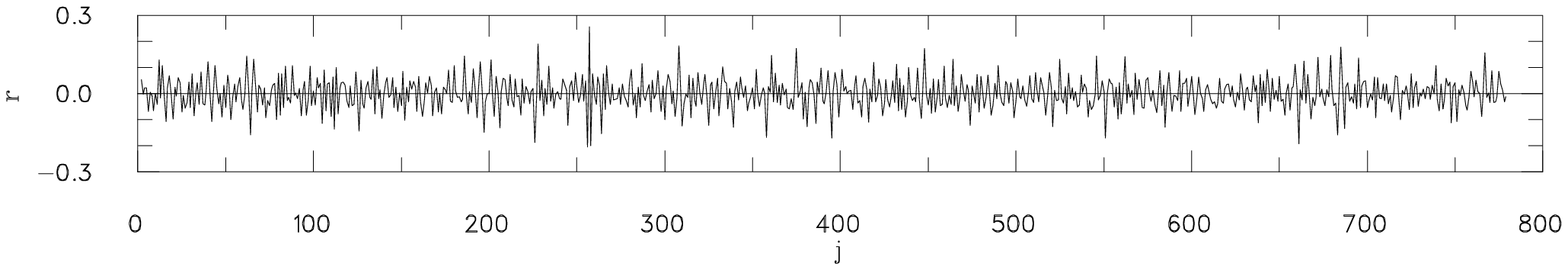}\\
\includegraphics[angle=90,width=134mm]{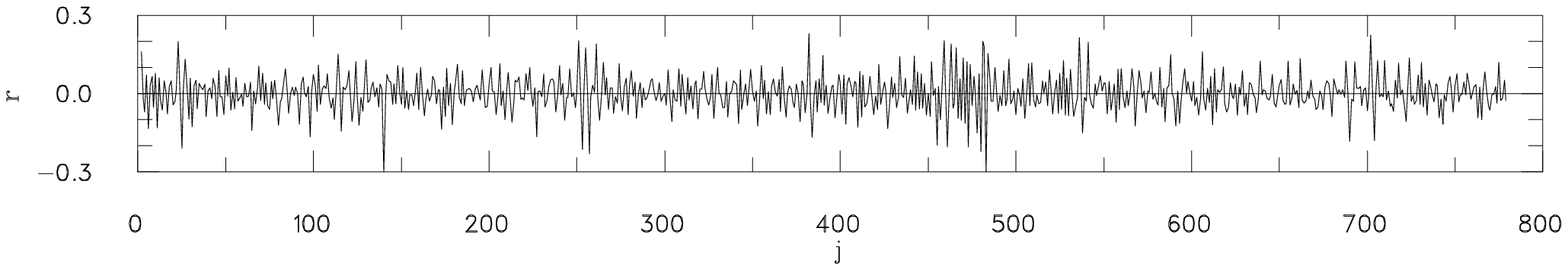}\\
\includegraphics[angle=90,width=134mm]{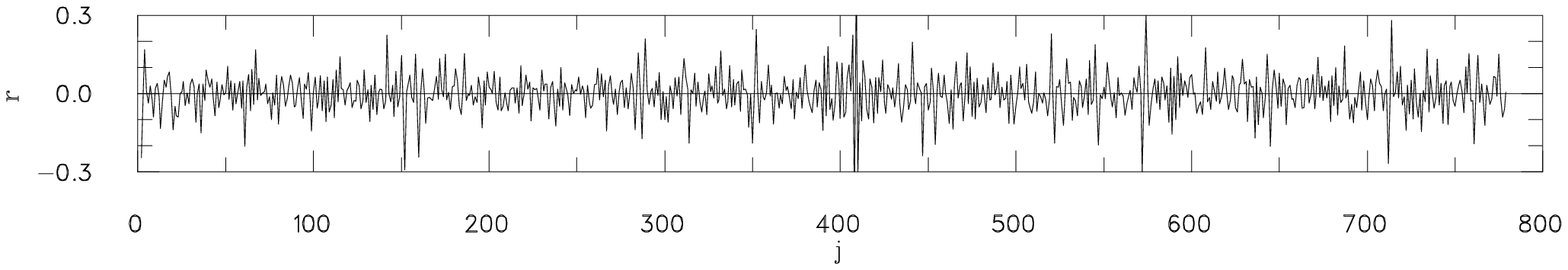}\\
\includegraphics[angle=90,width=134mm]{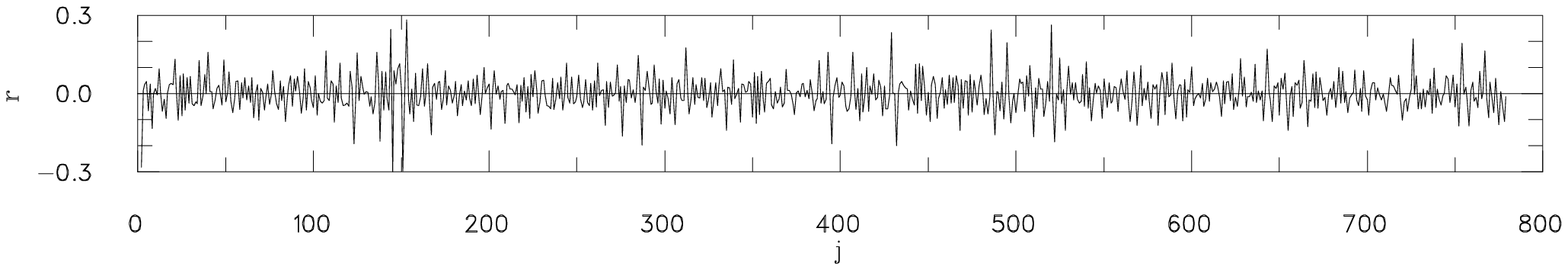}\\
\includegraphics[angle=90,width=134mm]{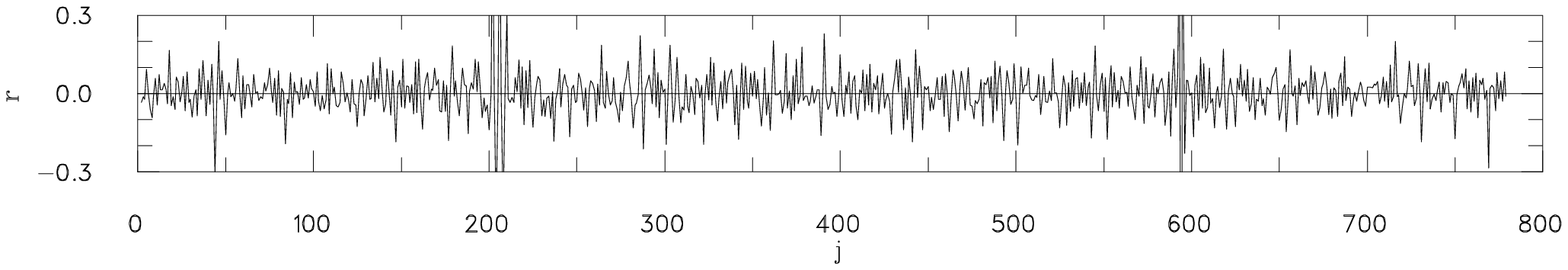}\\
\includegraphics[angle=90,width=134mm]{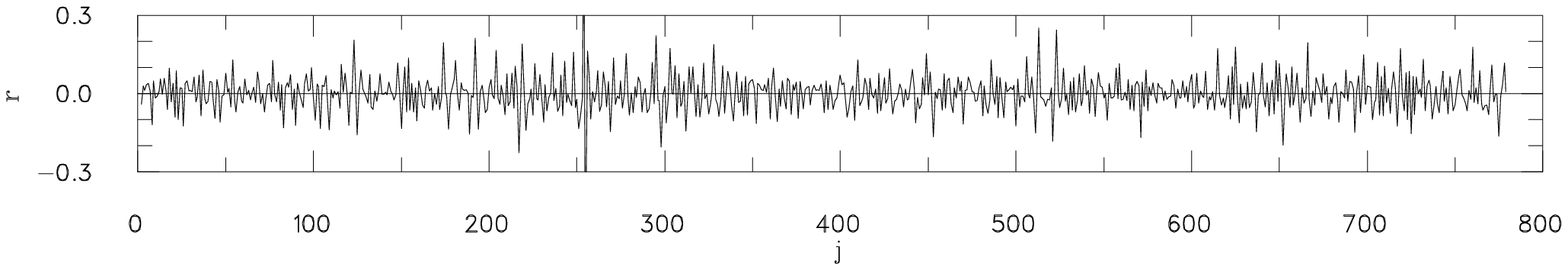}\\
\includegraphics[angle=90,width=134mm]{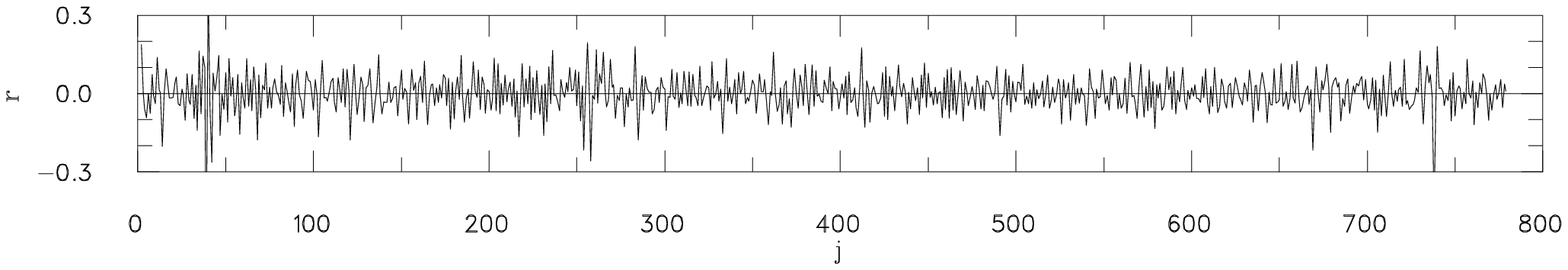}\\
\includegraphics[angle=90,width=134mm]{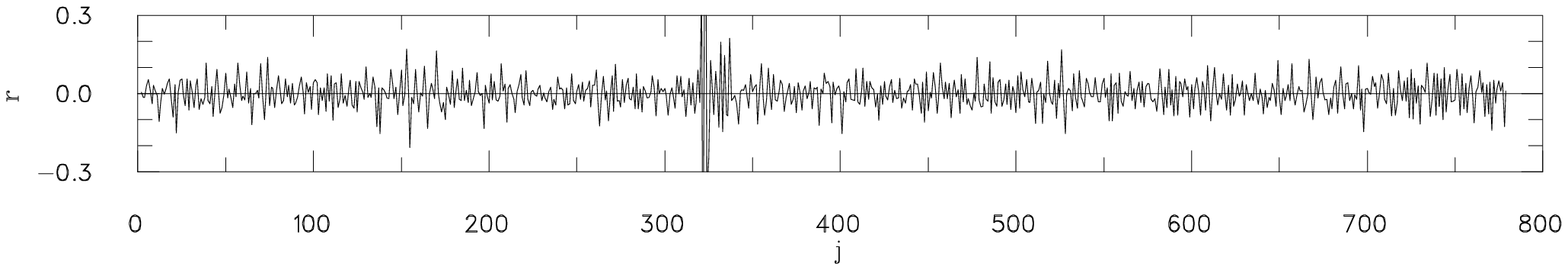}
\caption{\label{fig:tr13}Samples of returns time series for the update parameter
$\chi=0013$. The corresponding gains distribution is shown
in Fig.~\protect\ref{fig:gd}.}
\end{figure}

Samples of those time series are displayed in Fig.~\ref{fig:tr05} for $\chi=0.0005$
and Fig.~\ref{fig:tr13} for $\chi=0.0013$, respectively.
Each figure is composed of eight randomly selected independent simulation runs with
$4\times 10^6$ updates each.
Compared to the $\chi=0.0013$ time series, the $\chi=0.0005$ series have a,
somewhat perceptible, higher occurrence of
volatility clusters as well as bigger swings between them.

We now turn to gauging the ability of our model to replicate various 
features of financial markets.  One crucial aspect of financial markets returns is 
their volatility, as well as how this volatility evolves over time.  Volatility is 
more relevant today than ever, with large spikes possibly occurring in short periods 
of time. Financial markets returns generally display volatility `clusters'. These 
clusters indicate that once the volatility is high, it tends to remain high for a 
while, and that similarly, once it has come down, it tends to remain low for some 
time.  A convenient and well-accepted way of modeling such characteristics is through 
the use of the Auto Regressive Conditional Heteroskedasticity (ARCH) model pioneered 
by Engle \cite{Engle:1982:ARCH} or through the use of the more encompassing
Generalized Regressive Conditional Heteroskedasticity (GARCH) model proposed by
Bollerslev \cite{Bollerslev:1986:GARCH}.

Whether working on pricing a derivative product, attempting to hedge an exposure, optimizing a 
portfolio in a mean-variance framework, or estimating the Value-At-Risk of a position, 
the ability to capture and model the stochasticity and the clustering properties of 
the volatility is paramount. Not doing so can lead to the wrong probability 
distribution being used, since volatility clusters impact the shape of returns 
distributions in two important ways. First, the fact that a period of calm 
statistically tends to be followed by another period of calm indicates that there will 
be a fairly large amount of probability mass around the mean (return). Graphically 
this phenomenon translates into a probability distribution function that is higher 
than the Gaussian one in the vicinity of the mean.  Second, the fact that a period of 
extreme movements statistically tends to be followed by another period of extreme 
movements indicates that there will be non-negligible amounts of probability mass in 
the tails areas. Graphically this phenomenon translates into a probability 
distribution function that is higher than the Gaussian one in the vicinity of the 
tails.

The fairly wide-ranging GARCH specification models the volatility as both a 
function of past squared return shocks and of past levels of itself. If both past 
return shocks and past volatility levels are low, for instance, the odds are that the 
next volatility levels will remain low. If past volatility levels are low but recent 
past return shocks are high, the levels of volatility will likely increase. If past 
volatility levels are high but recent past return shocks are low, the levels of 
volatility will perhaps decrease. Finally, if both past return shocks and past 
volatility levels are high, the odds are that the next volatility levels will remain 
high. In a GARCH(p,q) specification, the number of lags $p$ and $q$ allowed for past 
shocks and past volatility levels are limitless. However, Hansen and
Lunde \cite{Lunde+Hansen:2005} explore the ability of 330 different ARCH/GARCH
models to capture the features of 
various financial returns and come to the conclusion that a GARCH(1,1) model performs 
just as well as the more `sophisticated' ones. Our specification for the
volatility $\sigma_t^2$ at time $t$ is thus a GARCH(1,1) described by the following
equation:
\begin{equation}
\sigma^2_t = \alpha_0+\alpha_1\epsilon_{t-1}^{2}+\beta_1\sigma_{t-1}^{2} \,,
\label{eq30}\end{equation}
where $\epsilon_{t-1}^{2}$ and $\sigma_{t-1}^{2}$ are the one-period lagged squared
return shock and one-period lagged variance, respectively.

We simulate 100 lattice time series for two different levels of our tuning 
parameter $\chi$. In the first set of simulations, $\chi$ is set equal to a value
of $0.0005$ while 
in our second set of simulations, $\chi$ is set equal to a value of $0.0013$.
This allows for 
the generation of different returns distributions and times series, namely financial 
markets returns with varying degrees of activity: The set of simulations using a 
$\chi$ value of $0.0005$ reflect a somewhat more volatile market than the set of simulations 
using a $\chi$ value of $0.0013$. Previous studies have shown that even a simple lattice 
model \cite{Dupoyet20113120} is able to produce returns volatility dynamics displaying
some ARCH/GARCH 
effects, and also that it is able to produce returns volatility dynamics that are 
rather consistent with those of NASDAQ historical returns. With the present model, we 
are here conducting a more extensive investigation: We estimate GARCH(1,1) fits from a 
large number of randomly chosen lattice configurations for the purpose of
demonstrating the stability, consistency, and flexibility of our lattice model.

\begin{figure}[!ht]
\rule{1.7mm}{0mm}\includegraphics[angle=90,width=128.0mm]{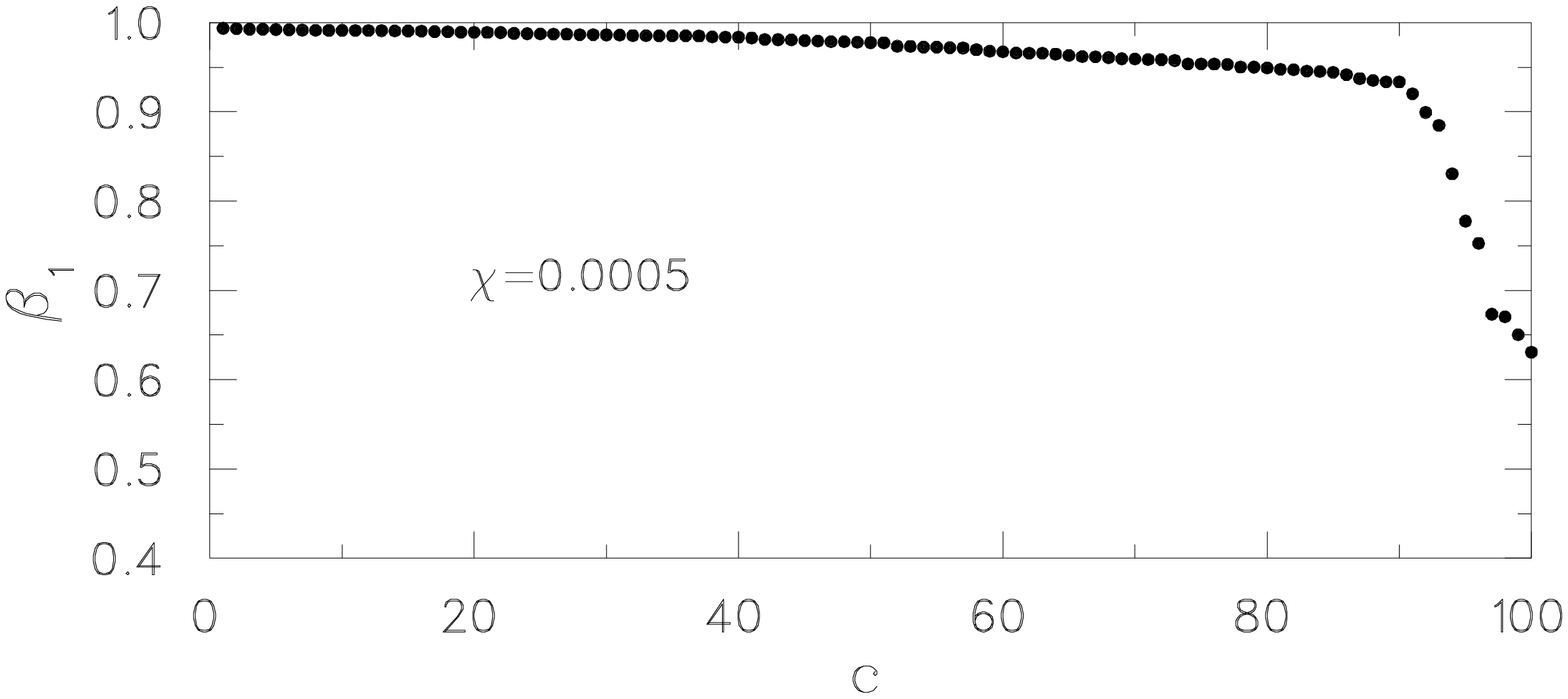}\vspace{3ex}\\
\rule{2.6mm}{0mm}\includegraphics[angle=90,width=127.2mm]{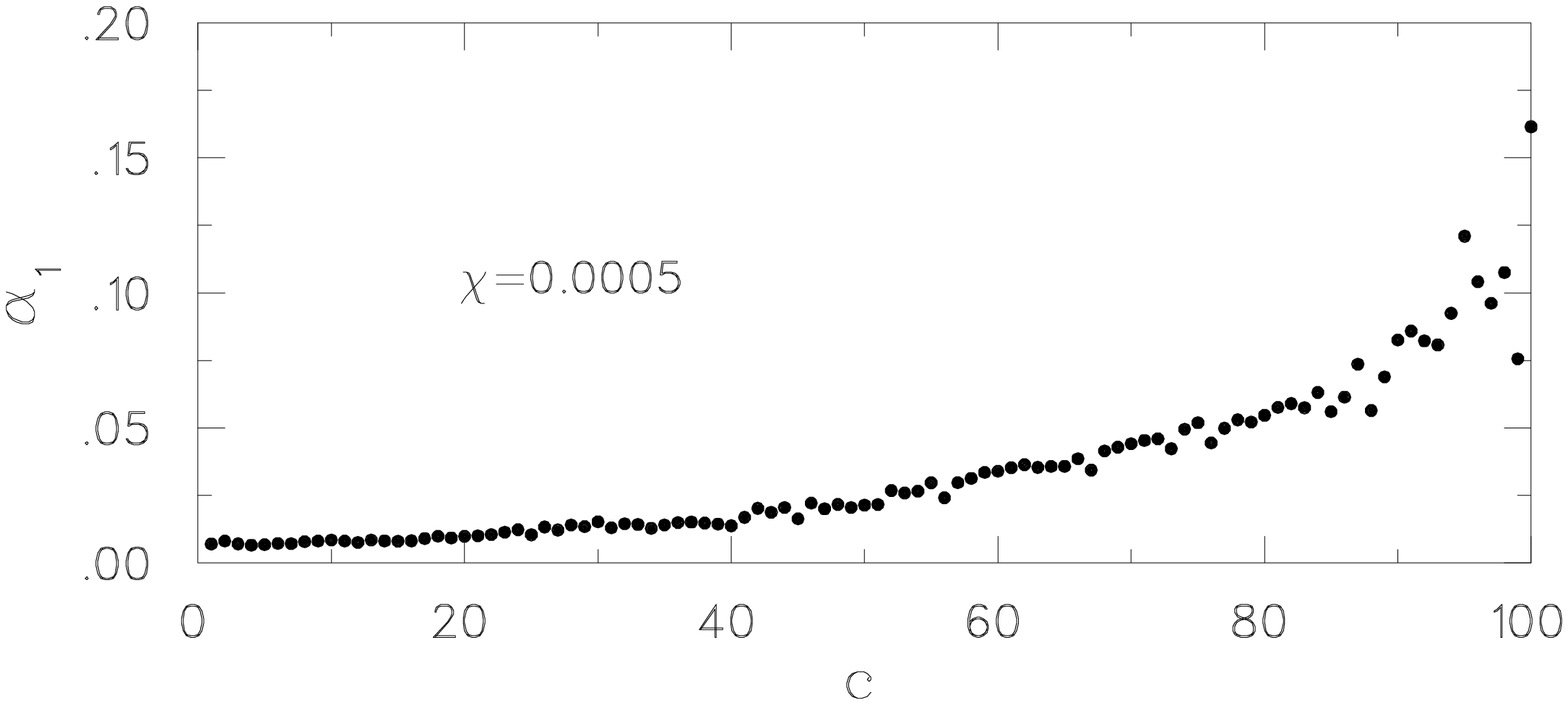}\vspace{3ex}\\
\includegraphics[angle=90,width=130.0mm]{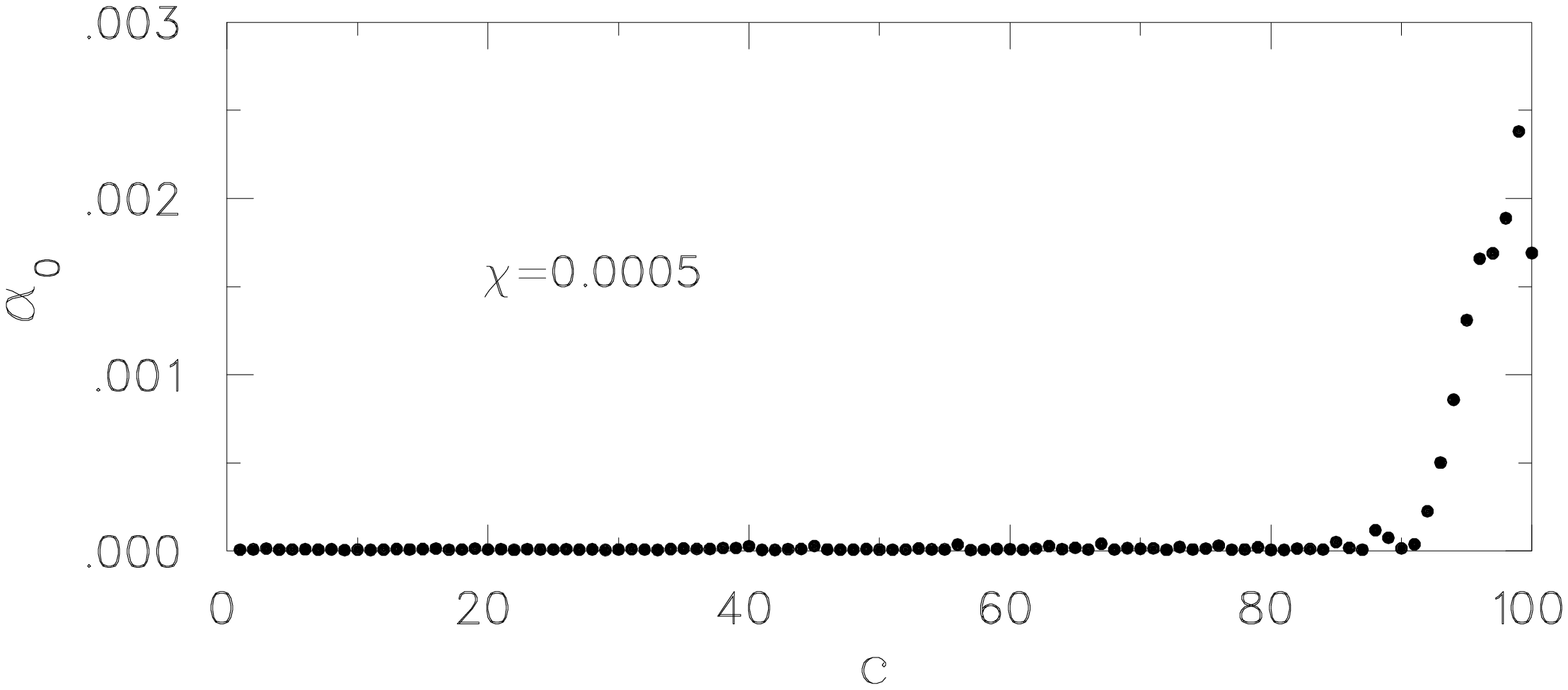}
\caption{\label{fig:ag0005}GARCH(1,1) fit parameters for a sample of 100 lattice time series
at $\chi=0.0005$. The configuration counter $c$ has been subject to sorting with respect to $\beta_1$.}
\end{figure}

\begin{figure}[!ht]
\rule{1.7mm}{0mm}\includegraphics[angle=90,width=128.0mm]{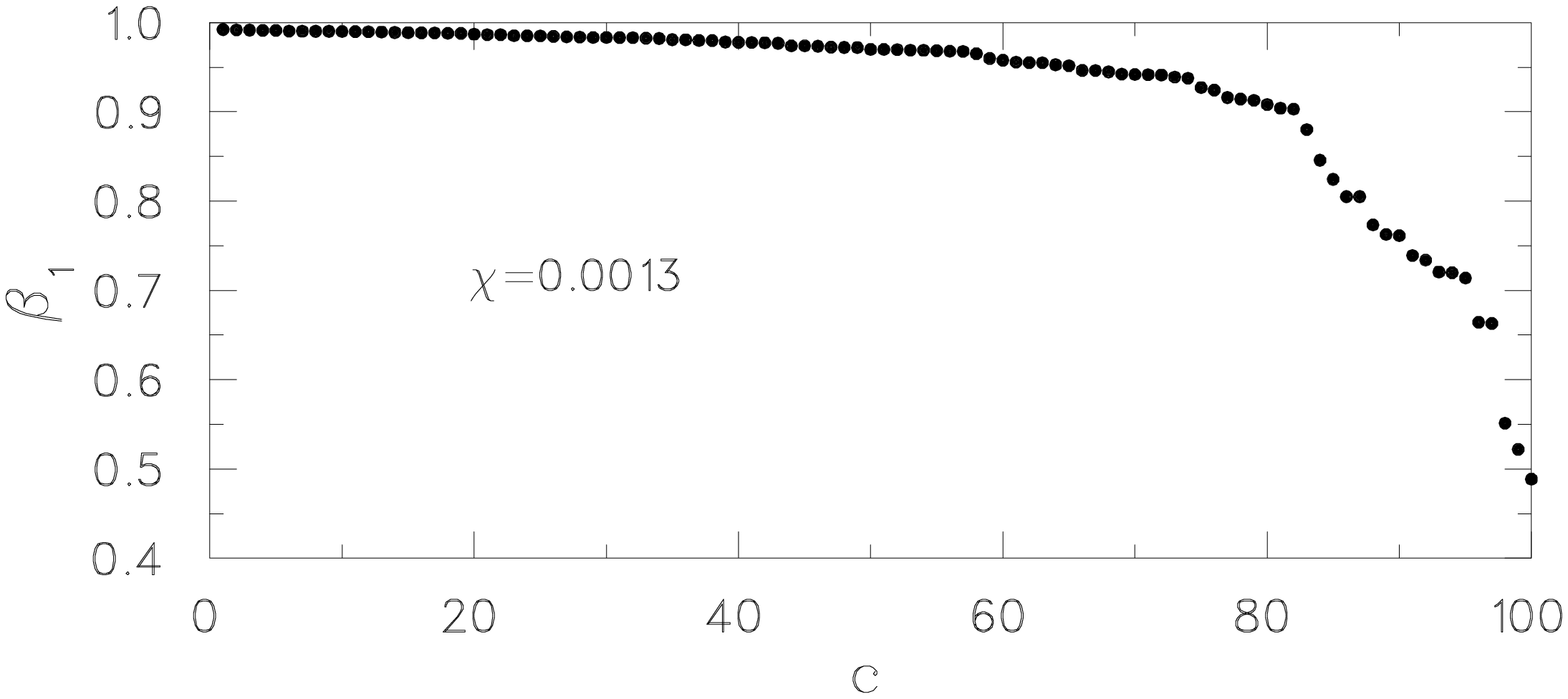}\vspace{3ex}\\
\rule{2.6mm}{0mm}\includegraphics[angle=90,width=127.2mm]{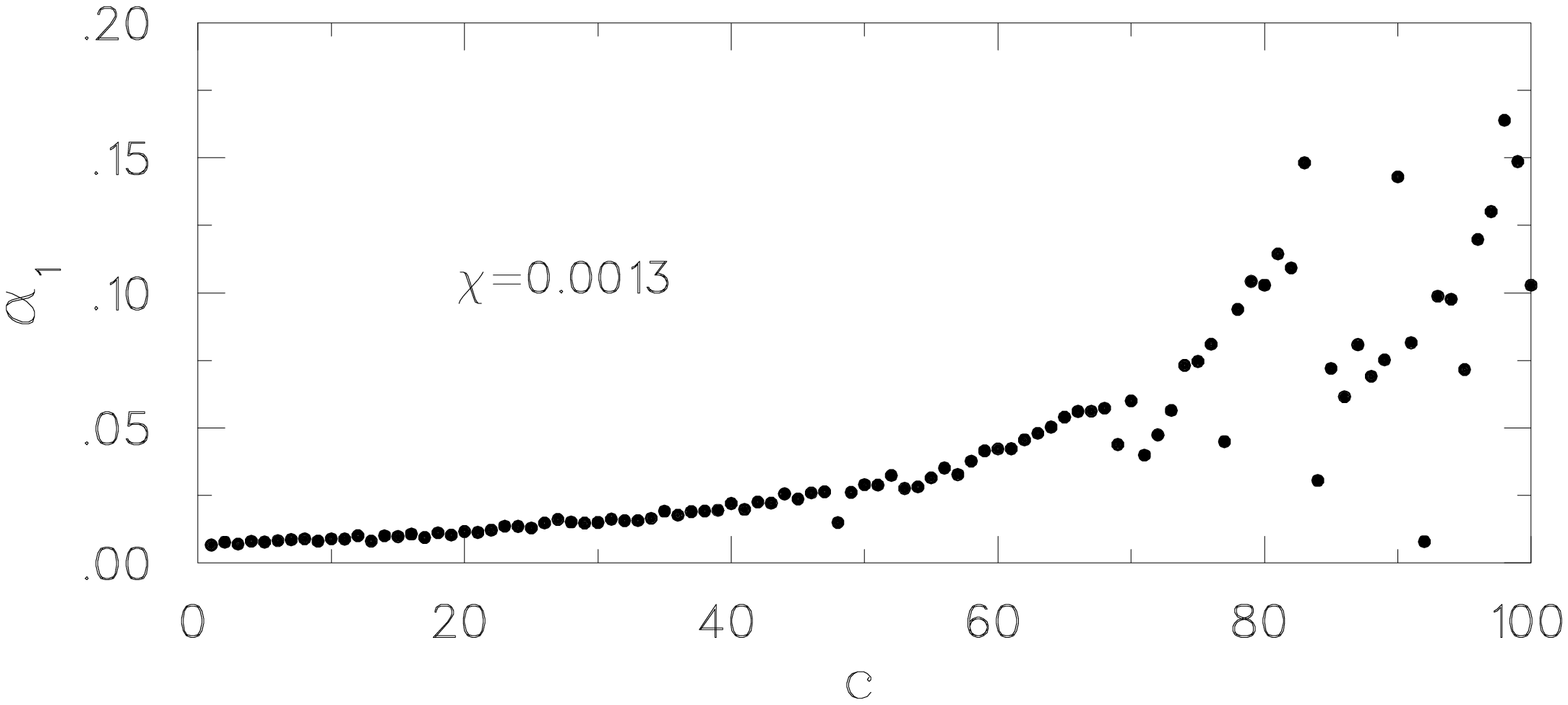}\vspace{3ex}\\
\includegraphics[angle=90,width=130.0mm]{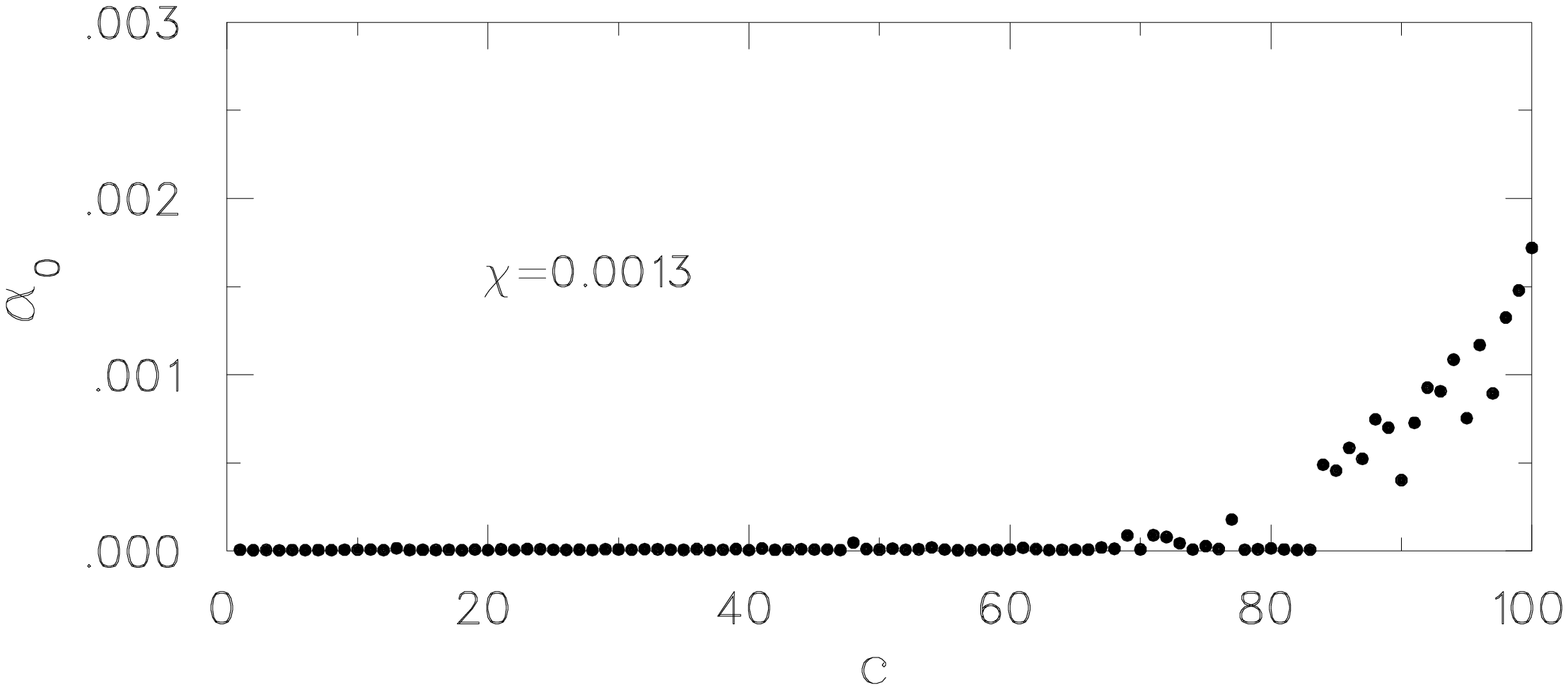}
\caption{\label{fig:ag0013}GARCH(1,1) fit parameters for a sample of 100 lattice time series
at $\chi=0.0013$. The configuration counter $c$ has been subject to sorting with respect to $\beta_1$.}
\end{figure}

The resulting parameters obtained 
are shown in Fig.~\ref{fig:ag0005} and in Fig.~\ref{fig:ag0013} for $\chi=0.0005$
and $\chi=0.0013$ respectively, for a total 
of 200 different lattice time series. For visual ease, the parameters are ordered 
before being plotted. In Fig.~\ref{fig:ag0005} one can thus immediately see that
the $\beta_1$ parameter 
is between $0.85$ and $1.0$ in about 95\% of the cases, a result consistent with Nasdaq 
GARCH(1,1)-estimated figures in \cite{Dupoyet20113120}.
In Fig.~\ref{fig:ag0013} the $\beta_1$ parameter is between $0.85$ and 
$1.0$ in about 80\% of the cases. This indicates that the lagged volatility feedback is 
strongly consistent across various market conditions and various simulations.
In Fig.~\ref{fig:ag0005} one can also see that the $\alpha_1$ parameter is
below $0.15$ in about 99\% of the cases, a 
result also consistent with Nasdaq GARCH(1,1)-estimated figures in \cite{Dupoyet20113120}.
Finally, in Fig.~\ref{fig:ag0013} the $\alpha_1$ parameter is below $0.15$ in about 99\% of
the cases, although a higher proportion is above $0.05$ when compared to the values it
takes in Fig.~\ref{fig:ag0005}. These 
results indicate that the lagged return shock feedback is also strongly consistent 
across various market conditions and various simulations. The $\alpha_0$ parameter is 
consistently very low in both cases, as it should at these high frequencies.

It is also interesting to note that for each parameter, we obtain some `outliers' in
about 5\% to 20\% of the cases, although they are outliers only in the sense that they differ 
from the other 95\% to 80\% of the estimates that are themselves incredibly consistent, 
and are not outliers in the sense that their values would be considered too extreme or 
unreasonable. It is important to note that when running multiple simulations, one is 
bound to obtain some results that differ somewhat from the estimates' consensus. For 
instance, even if one was to simulate an exact GARCH(1,1) process many times and 
subsequently estimate the parameters from the simulated data, some percentage of the 
estimated parameters would stray from the median estimates. In summary, 100\% of our 
estimated parameters are sensible, across various market conditions (more or less 
volatile) and thus across varying returns distributions, and over a myriad of 
simulations.

\section{\label{sec:conclusion}Conclusion}

The subject of this work has been to explore a class of models designed to 
simulate the properties of financial markets. The output of the model is a 
time series of returns, from which gains distributions and related features 
could be derived. The potential of the model for replicating market dynamics, 
as described by standard financial analysis tools, was the primary aim of 
this study.

The production of our time series has been done by numerical 
simulation based on a lattice description of fields in time-asset space. We 
have restricted ourselves to a one-asset model linked to an interest rate 
account. The dynamics are based on a gauge invariant lattice action which, 
when quantized, gives rise to eliminating arbitrage opportunities up to 
stochastic fluctuations, thus reflecting real market conditions. The second 
pillar of the model is an updating prescription that evolves the lattice 
fields into a self-organizing critical state. This appears to be an essential 
element for reproducing certain stylized features of real markets.

As a third 
feature, a parameter has been introduced as a tuning tool, through which a 
variety of market characteristics, ranging from quiescent to volatile 
markets, can be modeled.

An extensive analysis of a very large number of time 
series features evaluated by a GARCH(1,1) analysis was performed. It turns 
out that close to 100\% of the lattice model-generated time series give rise 
to sensible analysis parameters, rendering the model results almost 
indistinguishable from historical market data. In particular we could verify 
this observation across various market conditions and varying returns 
distributions.

We conclude that the model shows promise as a modeling tool 
for financial time series and look forward to further development and 
applications.

\appendix

\section{\label{sec:A}Gauge fixing}

In the notation of \cite{Dupoyet2010107}, the probability density for a particular
gauge field component has the form
\begin{equation}
p_\Theta(\theta_\mu(x)) \propto \exp(-\beta(\bar{L}_\Theta\exp(\theta_\mu(x))+\exp(-\theta_\mu(x))L_\Theta)\,)
\label{A1}\end{equation}
where $L_\Theta$ and $\bar{L}_\Theta$ are positive coefficients independent of $\theta_\mu(x)$.
They reflect the (local) environment of the link variable.
Under a gauge transformation, writing
\begin{equation}
g(x)=e^{h(x)}\,,
\label{A2}\end{equation}
we have
\begin{eqnarray}
\theta_\mu(x) &\rightarrow& \theta^\prime_\mu(x)=h(x)+\theta_\mu(x)-h(x+e_\mu) \label{A3} \\
L_\Theta &\rightarrow& L^\prime_\Theta=e^{h(x)}L_\Theta e^{-h(x+e_\mu)} \label{A4} \\
\bar{L}_\Theta &\rightarrow& \bar{L}^\prime_\Theta=e^{h(x+e_\mu)}\bar{L}_\Theta e^{-h(x)} \,. \label{A5}
\end{eqnarray}
The transformation laws for $L_\Theta$ and $\bar{L}_\Theta$
can be derived directly by an examination of the lattice action $S[\Theta,\Phi,\bar{\Phi}]$.
They are also obvious from the fact that the action is gauge invariant and the arguments
of the exponential functions of (\ref{A1}) are made up from invariant contributions to it. 
We now choose the gauge transformation by requiring the new coefficients to be equal
\begin{equation}
\frac{L^\prime_\Theta}{\bar{L}^\prime_\Theta} = e^{2h(x)}\frac{L_\Theta}{\bar{L}_\Theta} e^{-2h(x+e_\mu)}=1\,,
\label{A6}\end{equation}
or
\begin{equation}
h(x+e_\mu)-h(x) = \frac{1}{2}\log(\frac{L_\Theta}{\bar{L}_\Theta})\,.
\label{A7}\end{equation}
A solution of (\ref{A7}) is
\begin{eqnarray}
h(x+e_\mu) &=& \frac{\lambda+1}{4}\log(\frac{L_\Theta}{\bar{L}_\Theta}) \label{A8} \\
h(x) &=& \frac{\lambda-1}{4}\log(\frac{L_\Theta}{\bar{L}_\Theta}) \label{A9} \,,
\end{eqnarray}
where $\lambda$ is a real parameter\footnote{It may be used to control the
effect of the gauge transformation on the matter fields $\phi(x)$ and $\phi(x+e_\mu)$.}.
On all other sites, besides $x$ and $x+e_\mu$, the gauge transformation function $h(x)$
is arbitrary. (For definiteness one may choose it to be zero.)
In the new gauge, using (\ref{A6}), we have
\begin{equation}
\bar{L}^\prime_\Theta e^{\theta^\prime_\mu(x)}+e^{-\theta^\prime_\mu(x)}L^\prime_\Theta =
2\sqrt{L^\prime_\Theta\bar{L}^\prime_\Theta}\cosh(\theta^\prime_\mu(x)) \,.
\label{A10}\end{equation}
Thus, the probability distribution function is
\begin{equation}
p^\prime_\Theta(\theta^\prime_\mu(x)) \propto
\exp(-2\beta \sqrt{L^\prime_\Theta\bar{L}^\prime_\Theta} \cosh(\theta^\prime_\mu(x))\,)\,.
\label{A11}\end{equation}
Dropping the primes gives (\ref{eq18}).
Citing ${L_\Theta\bar{L}_\Theta}={L^\prime_\Theta\bar{L}^\prime_\Theta}$, we note
that the variance of the the probability distribution function is not altered by the
gauge transformation.

Again, in the notation of \cite{Dupoyet2010107}, the probability density for a particular
matter field component has the form
\begin{equation}
p_\Phi(\phi_\mu(x)) \propto \exp(-\beta(\bar{L}_\Phi\exp(\phi_\mu(x))+\exp(-\phi_\mu(x))L_\Phi)\,)
\label{A12}\end{equation}
where $L_\Phi$ and $\bar{L}_\Phi$ are positive coefficients independent of $\phi_\mu(x)$,
reflecting the (local) environment of the field variable.
In this case, changing the gauge (\ref{A2}) entails the transformations
\begin{eqnarray}
\phi(x) &\rightarrow& \phi^\prime(x)=h(x)+\phi(x) \label{A13} \\
L_\Phi &\rightarrow& L^\prime_\Phi=e^{h(x)}L_\Phi \label{A14} \\
\bar{L}_\Phi &\rightarrow& \bar{L}^\prime_\Phi=\bar{L}_\Phi e^{-h(x)} \,. \label{A15}
\end{eqnarray}
The requirement $L^\prime_\Phi/\bar{L}^\prime_\Phi=1$ leads to
\begin{equation}
h(x) = -\frac{1}{2}\log(\frac{L_\Phi}{\bar{L}_\Phi})\,,
\label{A16}\end{equation}
while the gauge function is arbitrary on all sites other than $x$.
Proceeding in the manner above we have
\begin{equation}
\bar{L}^\prime_\Phi e^{\phi^\prime_\mu(x)}+e^{-\phi^\prime_\mu(x)}L^\prime_\Phi =
2\sqrt{L^\prime_\Phi\bar{L}^\prime_\Phi}\cosh(\phi^\prime_\mu(x)) \,,
\label{A17}\end{equation}
and thus obtain the probability distribution function for the matter field
\begin{equation}
p^\prime_\Phi(\phi^\prime_\mu(x)) \propto
\exp(-2\beta\sqrt{L^\prime_\Phi\bar{L}^\prime_\Phi}\cosh(\phi^\prime_\mu(x))\,)\,.
\label{A18}\end{equation}
Dropping the primes gives (\ref{eq19}). Again, because of 
${L_\Phi\bar{L}_\Phi}={L^\prime_\Phi\bar{L}^\prime_\Phi}$, the gauge
transformation does not change the variance
of the distribution.

\end{document}